\documentclass[
aip,
pop,
reprint,
]{revtex4-1}
\usepackage{graphicx}
\usepackage{amsmath}
\usepackage{amssymb}
\usepackage{bm}
\usepackage{natbib}

\newcommand{\grad}{\boldsymbol{\nabla}}

\newcommand{\ue}{\ensuremath{\mathbf{u}_{_E}}}
\newcommand{\vv}[1]{\ensuremath{\mathbf{#1}}}

\DeclareMathOperator{\sech}{sech}

\begin{document}

%% LaTeX will automatically break titles if they run longer than
%% one line. However, you may use \\ to force a line break if
%% you desire.

\title[Electron Heating and Acceleration during Reconnection]{The Mechanisms of Electron Heating and Acceleration during Magnetic Reconnection}
%\thanks{Footnote to title of article.}

\author{J. T. Dahlin}
 \email{jdahlin@umd.edu}
 \affiliation{Institute for Research in Electronics and Applied Physics, University of Maryland, College Park, MD 20742}
\author{J. F. Drake}
 \affiliation{Institute for Research in Electronics and Applied Physics, University of Maryland, College Park, MD 20742}
 \affiliation{Department of Physics, University of Maryland, College Park, MD 20742}
 \affiliation{Institute for Physical Science and Technology, University of Maryland, College Park, MD 20742}
 \affiliation{Space Science Laboratory, University of California, Berkeley, CA 94720}
\author{M. Swisdak}
 \affiliation{Institute for Research in Electronics and Applied Physics, University of Maryland, College Park, MD 20742}
\date{\today}

\begin{abstract}

The heating of electrons in collisionless magnetic reconnection is
explored in particle-in-cell (PIC) simulations with non-zero guide
fields so that electrons remain magnetized.  In this regime electric
fields parallel to $\mathbf{B}$ accelerate particles directly while
those perpendicular to $\mathbf{B}$ do so through gradient-$B$ and
curvature drifts.  The curvature drift drives parallel heating through
Fermi reflection while the gradient $B$ drift changes the
perpendicular energy through betatron acceleration. We present
simulations in which we evaluate each of these mechanisms in space and
time in order to quantify their role in electron heating. For a case
with a small guide field (20 \% of the magnitude of the reconnecting
component) the curvature drift is the dominant source of electron
heating.  However, for a larger guide field (equal to the magnitude of
the reconnecting component) electron acceleration by the curvature
drift is comparable to that of the parallel electric field. In both
cases the heating by the gradient $B$ drift is negligible in
magnitude. It produces net cooling because the conservation of the
magnetic moment and the drop of $B$ during reconnection produce a
decrease in the perpendicular electron energy. Heating by the
curvature-drift dominates in the outflow exhausts where bent field
lines expand to relax their tension and is therefore distributed over a
large area. In contrast, the parallel electric field is localized near
X-lines. This suggests that acceleration by parallel electric fields
may play a smaller role in large systems where the X-line occupies a
vanishing fraction of the system. The curvature drift and the parallel
electric field dominate the dynamics and drive parallel heating. 
A consequence is that the electron energy spectrum becomes 
extremely anisotropic at late time, which has important
 implications for quantifying the limits of electron acceleration 
 due to synchrotron emission. An upper limit on electron energy 
gain that is substantially higher than earlier estimates is obtained by balancing reconnection drive with radiative loss.

\end{abstract}

%% Keywords should appear after the \end{abstract} command. The uncommented
%% example has been keyed in ApJ style. See the instructions to authors
%% for the journal to which you are submitting your paper to determine
%% what keyword punctuation is appropriate.

%\keywords{KEY WORDS}

\maketitle

%% From the front matter, we move on to the body of the paper.
%% In the first two sections, notice the use of the natbib \citep
%% and \citet commands to identify citations.  The citations are
%% tied to the reference list via symbolic KEYs. The KEY corresponds
%% to the KEY in the \bibitem in the reference list below. We have
%% chosen the first three characters of the first author's name plus
%% the last two numeral of the year of publication as our KEY for
%% each reference.

%% Authors who wish to have the most important objects in their paper
%% linked in the electronic edition to a data center may do so by tagging
%% their objects with \objectname{} or \object{}.  Each macro takes the
%% object name as its required argument. The optional, square-bracket 
%% argument should be used in cases where the data center identification
%% differs from what is to be printed in the paper.  The text appearing 
%% in curly braces is what will appear in print in the published paper. 
%% If the object name is recognized by the data centers, it will be linked
%% in the electronic edition to the object data available at the data centers  
%%
%% Note that for sources with brackets in their names, e.g. [WEG2004] 14h-090,
%% the brackets must be escaped with backslashes when used in the first
%% square-bracket argument, for instance, \object[\[WEG2004\] 14h-090]{90}).
%%  Otherwise, LaTeX will issue an error. 

\section{Introduction}
\label{sec:introduction}

Magnetic reconnection is a ubiquitous plasma process that converts
magnetic energy into thermal and kinetic energy. Of particular
interest is the production of non-thermal particles, in which a
fraction of the plasma population is driven to energies much larger
than that found in the ambient medium. Reconnection is thought to be an
important driver of such particles in phenomena such as gamma ray bursts
\citep{drenkhahn02a,michel94a}, astral and solar flares
\citep{lin03a}, and magnetospheric storms \citep{oieroset02a}.  Recent
observations of solar flares reveal the remarkable efficiency of
electron acceleration: a large fraction of the electrons in the flaring
region become a part of the nonthermal spectrum, with a resulting
energy content comparable to that of the magnetic field
\citep{krucker10a,oka13a}.

Mechanisms for particle acceleration have been explored and compared
in a variety of papers,
e.g. \cite{hoshino01a,zenitani01a,drake05a,pritchett06a,egedal09a} and
\cite{oka10a}. Previous authors
\citep{litvinenko96a,drake05a,egedal12a} examined acceleration by
electric fields parallel to the local magnetic field. \cite{drake06a} proposed a mechanism whereby charged particles gain
energy as they reflect from the ends of contracting magnetic islands,
a process analogous to first-order Fermi acceleration of cosmic
rays. Similar energy gain takes place during the merging of magnetic
islands \citep{oka10a,drake10a}. In reconnection in either
collisionless \citep{drake06b,daughton11a} or weakly collisional
systems \citep{biskamp86a,loureiro07a,cassak09a,huang10a}, current
sheets break up into many magnetic islands , producing an ideal
environment for this mechanism.  Although such work has clearly shown
that non-thermal tails can be produced in simulations of magnetic
reconnection, which process or processes are the dominant drivers of
these tails remains an open question.

%[Several unresolved questions persist in studies of particle %acceleration.] 

Resistive MHD studies showed that thermal test particles injected into
simulations quickly developed non-thermal features, most notably
power-law tails, whose energy content was large
\citep{dmitruk04a,onofri06a}.  So, although kinetic simulations incur
a much greater computational expense, they are necessary for studies
of particle acceleration where the feedback of particles on the
accelerating field is important.  The current largest kinetic
simulations of reconnection have domain sizes of several hundred ion inertial
lengths ($d_i = c/\omega_{pi}$, where $\omega_{pi} = \sqrt{4\pi n_0
  e^2/m_i}$ and $n_0$ is the density) in two dimensions.
Three-dimensional simulations are more computationally intensive and
thus smaller. By comparison, a typical coronal density of
$10^9/\text{cm}^3$ implies $d_i \sim 10^3$ cm, in a system with a
typical scale of $10^9$ cm. A model capable of explaining observations
such as that of \cite{krucker10a} will involve a significant extrapolation to
physically relevant domains.

Astrophysical magnetic reconnection occurs in a wide variety of
regimes. The ``guide field" (magnetic field component perpendicular to
the reconnecting components and parallel to the reconnection electric
field) is an important parameter: anti-parallel (small guide field)
reconnection is common in the magnetotail, while component
reconnection is common in environments such as the solar corona, where
the guide field may be many be several times the magnitude of the
reconnecting field.  Anti-parallel reconnection necessarily contains
regions where $|\mathbf{B}|$ is small and particles become
unmagnetized (i.e., Larmor radii exceed local scale lengths).
Electrons remain magnetized during guide field reconnection even for
very small values of the ratio of the guide field to the reconnecting
field \citep{swisdak05a}.

Electrons and ions, having disparate masses, can
be accelerated by distinct mechanisms. Small-scale processes
important for electrons may be washed out at the relatively large ion
length scales. Acceleration arising from discontinuities may have a
greater impact on massive particles. For example, \cite{knizhnik11a}
and \cite{drake14a} explore how heavy ions are
accelerated by a pick-up process in reconnection outflows, a mechanism
which does not impact electrons.  We consider only electron
acceleration in this paper.

Fully three-dimensional treatments are an ultimate goal in the kinetic
treatment of particle acceleration. In two dimensions, magnetic
islands (closed loops of magnetic flux) are efficient particle
traps. However, oblique tearing modes and other instabilities in 3D
generate flux tubes \citep{daughton11a}. These three-dimensional
equivalents of islands are porous, allowing particles to escape. 2D
models dependent on the structure of the magnetic fields (e.g. the
contracting islands model) may require modification in order to
address acceleration in the more complex geometrical situation in
3D. Hence a more general description of how particles are accelerated
during reconnection in 2D and 3D systems is needed.

In order to more fully establish the mechanisms for electron acceleration 
during reconnection, we develop a  guiding-center theory that can be 
used to diagnose how electrons gain energy during reconnection. We 
identify key mechanisms, including the curvature, gradient B and 
parallel electric fields,  and evaluate their relative contributions 
in 2D kinetic simulations with a modest to strong guide field. The 
results are easily generalizable to 3-D configurations. This is a 
similar approach to that used by \cite{guo14a} in their treatment of 
particle acceleration in relativistic, antiparallel reconnection.

In Section \ref{sec:theoretical} we derive a local bulk expression for
particle acceleration and discuss the physical significance of each
term. In Section \ref{sec:simulation} we delineate our simulation
methods and initial conditions. We present the results from 2D
simulations in Section \ref{sec:results} and examine momentum spectra in Section \ref{sec:spectra}. We discuss the significance
of these results in Section \ref{sec:conclusions}.

\section{Electron acceleration in the guiding center limit}
\label{sec:theoretical}

In order to examine various effects contributing to the particle
energy evolution, we begin with a standard treatment of the
guiding-center approximation given by \cite{northrop63a}. The
evolution of the energy $\epsilon$ of a single particle in the
guiding-center limit is given by:

\begin{equation}
\label{eqn:particle}
 \frac{d \epsilon}{d t} = (\mu/\gamma ) \partial_t B + q(v_\parallel\mathbf{b}
+ \mathbf{v}_c+\mathbf{v}_g) \cdot \mathbf{E}
\end{equation}
where $\mathbf{b} = \mathbf{B}/|\mathbf{B}|$, $\mu = m\gamma^2v_\perp^2/2B$, is the magnetic moment, $\gamma$ is the relativistic Lorentz factor, $v_\parallel = \mathbf{v}
\cdot \mathbf{b}$, and $\mathbf{v}_c$ and $\mathbf{v}_g$ are
the curvature and grad-B drifts:
\begin{equation}
\label{eqn:curv}
\mathbf{v}_c = \frac{v_\parallel^2 \mathbf{b}}{\Omega_{ce}} \times \boldsymbol{\kappa}
\end{equation}
\begin{equation}
\label{eqn:gradb}
\mathbf{v}_g =  \frac{v_\perp^2 \mathbf{b}}{2 \Omega_{ce}} \times \frac{\nabla B}{B}
\end{equation}
In Eqs.~ (\ref{eqn:curv}) and (\ref{eqn:gradb}) the electron
cyclotron frequency $\Omega_{ce} = eB/\gamma m_ec$. The curvature is
$\boldsymbol{\kappa} = \mathbf{b} \cdot \grad \mathbf{b}$.  If we
sum over all particles in a local region,
(\ref{eqn:particle}) becomes:

\begin{equation}
\frac{d U}{d t} = \frac{\beta_\perp}{2} \frac{\partial}{\partial t}
\left(\frac{B^2}{8 \pi} \right) + \vv{E} \cdot \left[ \vv{J}_\parallel
  + \frac{\beta_\parallel}{2} \frac{c}{4\pi} \mathbf{B} \times
  \boldsymbol{\kappa} + \frac{\beta_\perp}{2} \frac{c}{4 \pi} \vv{B}
  \times \frac{\nabla B}{B} \right]
\end{equation}
which may be rewritten as:

\begin{equation}
\label{eqn:tot_heat}
\frac{d U}{d t}
= E_\parallel J_\parallel 
+ \frac{p_\perp}{B} \left( \frac{\partial B }{\partial t} + \ue \cdot \grad B \right) 
+ p_\parallel \ue \cdot \boldsymbol{\kappa}
\end{equation}
where $U$ is the total kinetic energy, $u_E$ is the E-cross-B drift, $p_\parallel$ is the parallel
pressure, $p_\perp$ the perpendicular pressure,
$\beta_\parallel = 8 \pi p_\parallel/B^2$, and $\beta_\perp = 8 \pi
p_\perp/B^2$.

\begin{figure}
%\epsscale{.80}
\includegraphics{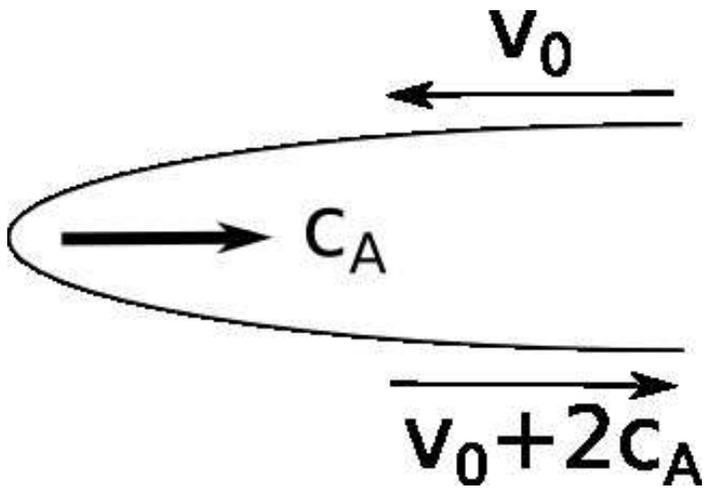}
\caption{Cartoon diagram of a charged particle reflecting from a
  magnetic loop contracting at the Alfv\'{e}n speed. The particle
  velocity increases by $2 c_A$.  }
\label{fermicartoon}
\end{figure}

The first term in Eq.~(5) is the acceleration by the parallel electric
field. The second term corresponds to perpendicular heating due to the
conservation of $\mu$ (the term in parentheses is $dB/dt$). The
third term drives parallel acceleration and arises from the
first-order Fermi mechanism described in \cite{drake06a,drake10a}.
Freshly reconnected field lines downstream from a reconnecting X-line
accelerate as a result of the tension force ($\sim
B^2\boldsymbol{\kappa}$) and causes them to ``straighten''. Particles
that reflect from this moving field line receive a Fermi ``kick'' and
thereby gain energy.  Figure \ref{fermicartoon} shows a cartoon model of
this effect. A particle trapped in a magnetic island whose ends are
contracting due to these tension forces will repeatedly gain energy as
it reflects from the ends of the island.  More generally, any particle
reflecting from this moving bent field line will gain energy. 
%The energy gain associated with such Fermi reflection is %proportional to  $\ue
%\cdot \boldsymbol{\kappa}$ in Eq.~(\ref{eqn:tot_heat}).

Though we do not address this directly in the present paper, we note
that the tension force is important for other aspects of energy
conversion. For example, it drives bulk ion outflow in the
reconnection exhaust and generates the counterstreaming ion distributions
that have been measured in the magnetosphere and solar wind
\citep{hoshino98a,gosling05a,phan07a}.

\section{Simulations}
\label{sec:simulation}

We explore particle heating via simulations using the particle-in-cell
(PIC) code {\tt p3d} \citep{zeiler02a}.  Particle trajectories are
calculated using the relativistic Newton-Lorentz equations, and the
electromagnetic fields are advanced using Maxwell's equations.

The initial condition consist of a uniform guide field $B_z$ superimposed on a
double-Harris equilibrium \citep{harris62a}. The magnetic field
configuration is:
\begin{equation}
   B_x = B_0\left[ \tanh\left(\frac{y-L_y/4}{w_0}\right)
       - \tanh\left(\frac{y-3L_y/4}{w_0}\right)-1 \right]
\end{equation}
where $B_0$ is the asymptotic reconnecting field, $w_0$ is the
current sheet half-width and $L_y$ is
the length of the computational domain in the $y$-direction.
The density consists of two populations, a drifting population 
with density:
\begin{equation}
   n = n_0 \left[ \sech^2 \left(\frac{y-L_y/4}{w_0}\right)
     + \sech^2 \left(\frac{y-3L_y/4}{w_0}\right) \right]
\end{equation}
that carries the current and a uniform background with density $0.2
n_0$. 

We use an artificial mass ratio $m_p/m_e=25$ and speed of light $c =
15 c_A$ where $m_p$ and $m_e$ are the electron mass, $c_A$ is the
Alfv\'{e}n velocity based on $B_0$ and $n_0$. These choices allow for
sufficient separation of scales (between proton and electron spatial
scales and electromagnetic and particle time scales, respectively)
while significantly reducing the computational expense of the
simulation.  Lengths in our simulation are normalized to the ion skin
depth $d_i = c/\omega_{pi}$ and times are normalized to the inverse
ion cyclotron frequency, $\Omega_{ci}^{-1}$. The initial temperature
of all species is $0.25m_ic_A^2$ for both background and
the current sheet populations. The current sheet half-thickness is set
to $w_0 = 0.25 d_i$ so that reconnection will onset quickly. The grid
scale is $\Delta = d_e/4 \approx 0.94 \lambda_D$ where $d_e =
c/\omega_{pe}$ is the electron inertial length and $\lambda_D$ is the
Debye length. We use periodic boundary conditions in both directions.

The goal of the present paper is to explore the mechanisms for
particle acceleration using the expression for electron energy gain in
Eq.~(\ref{eqn:tot_heat}). Since this equation is valid only for
adiabatic motion, we limit our computations to systems with a non-zero
initial guide field. It was shown previously that electrons are
magnetized in reconnecting systems with guide field exceeding $0.1B_0$
\citep{swisdak05a}. In this paper we therefore focus on two simulations
with non-zero guide fields both with dimensions $L_x \times L_y =
204.8 \times 102.4$. Simulation `A' has $B_z = 0.2B_0$ and `B' has
$B_z = 1.0B_0$. We note that although our simulations contain two
current sheets, we will often present results from only the upper
current sheet. In all cases the other sheet exhibits similar
behavior. We will then present results from a much larger simulation
with dimensions $L_x \times L_y = 819.2 \times 409.6$ with a guide
field $B_z=1.0B_0$.

\section{Simulation results: electron heating}
\label{sec:results}

\begin{figure}
%\epsscale{.80}
\includegraphics{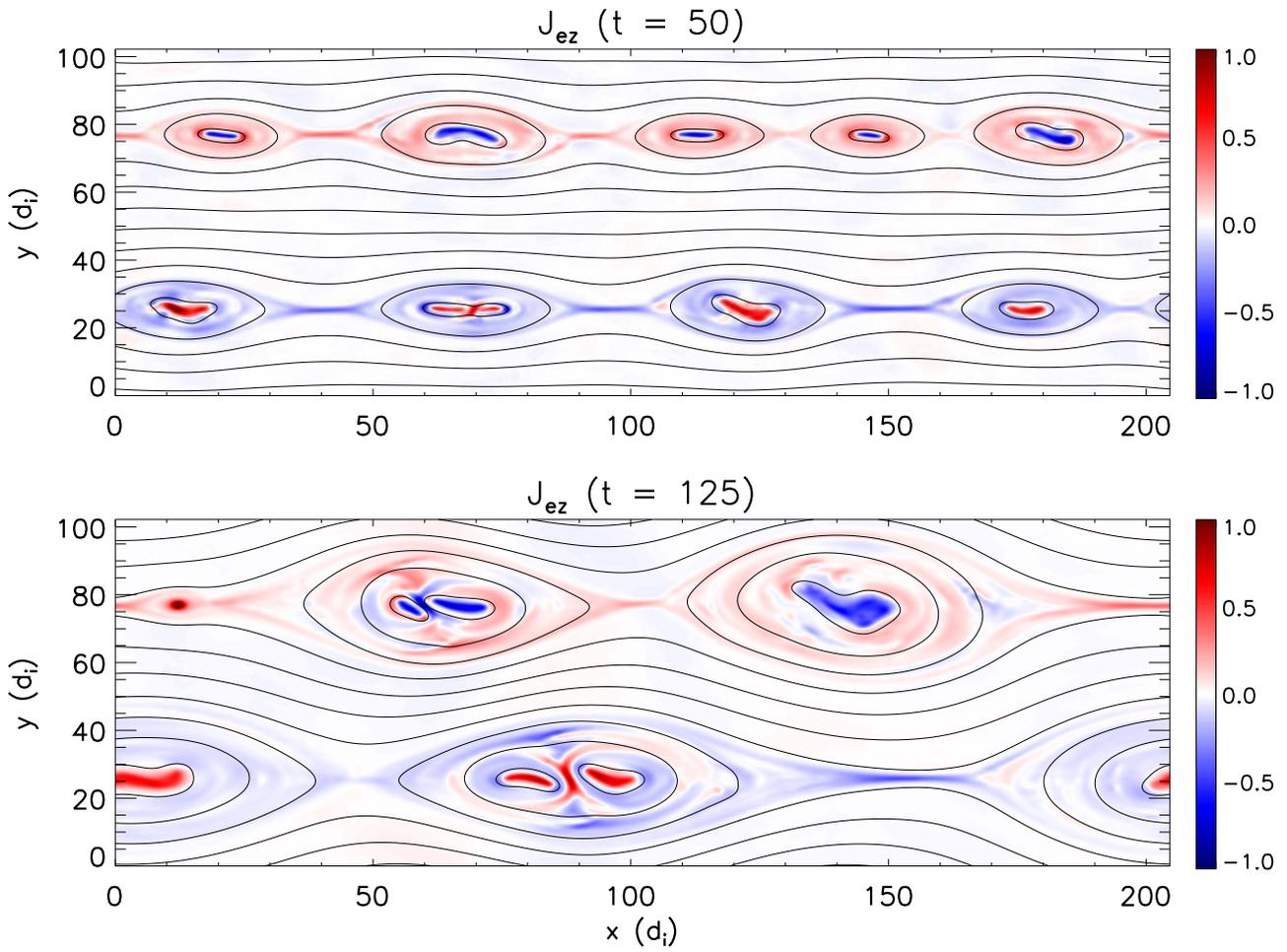}
\caption{Out-of plane electron current density in simulation A at $t
  \Omega_{ci} = 50$ (top) and $t \Omega_{ci} = 125$
  (bottom). Reconnection generates many islands which merge until they
  approach system the size.}
\label{jez}
\end{figure}

\begin{figure}
%\epsscale{.80}
\includegraphics{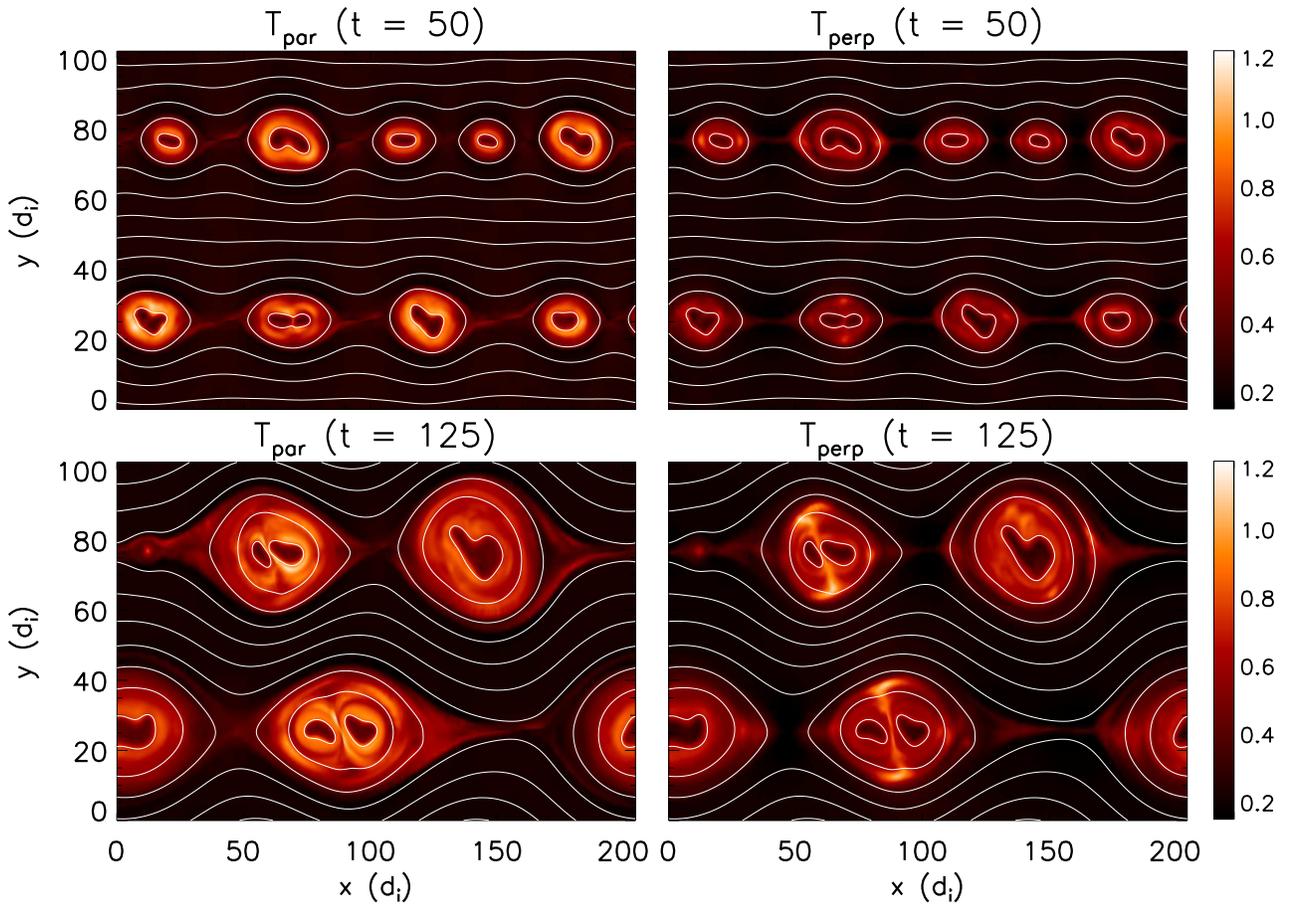}
\caption{Parallel and perpendicular electron temperature from a
  simulation with a guide field of $0.2$ (simulation A) at $t
  \Omega_{ci} = 50$ (top) and $t \Omega_{ci} = 125$ (bottom).  }
\label{teparperp}
\end{figure}

\begin{figure}
%\epsscale{.80}
\includegraphics{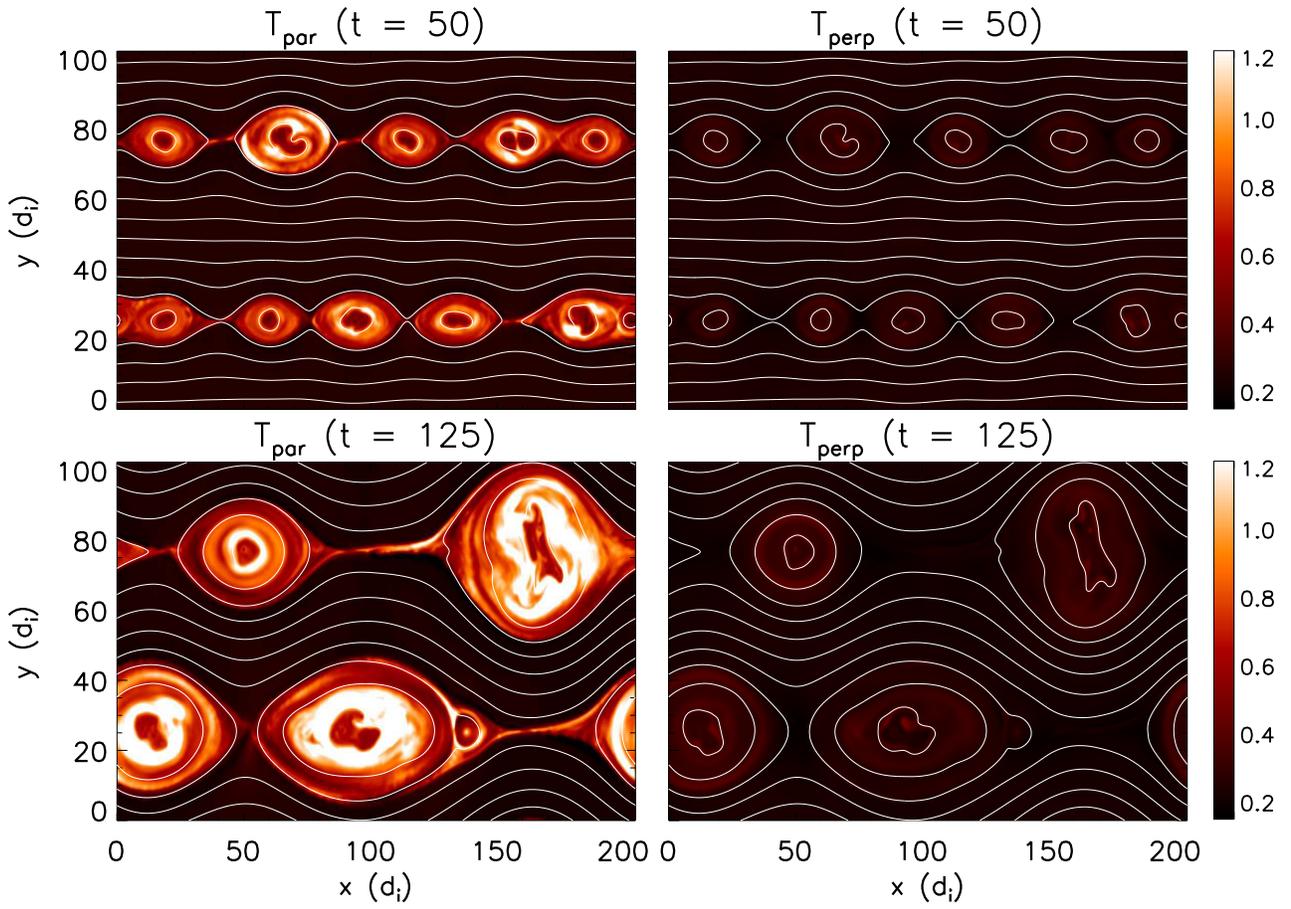}
\caption{Parallel and perpendicular temperatures from a simulation
  with a guide field of $1.0B_0$ (simulation B) at $t \Omega_{ci} =
  50$ (top) and $t \Omega_{ci} = 125$ (bottom).  }
\label{teparperp1}
\end{figure}

Reconnection develops rapidly from the particle noise inherent in the
PIC formulation. Figure \ref{jez} shows the evolution of the electron
out-of-plane current density in simulation A. The tearing instability
quickly generates many magnetic islands on each current layer which
continually grow and merge due to reconnection. We halt the simulation
when the islands on the two current layers begin to interact, here at
$t \approx 150$. Figures \ref{teparperp} and \ref{teparperp1} display
the parallel and perpendicular electron temperatures in simulations A
and B early and late in the simulation.  In simulation A the parallel
electron temperature increases substantially within the exhausts
downstream of the X-lines and within the developing magnetic
islands. The perpendicular temperature increment is strongest in
localized regions in the cores of magnetic islands.  In contrast,
$T_{e\parallel}$ significantly exceeds $T_{e\perp}$ throughout the
duration of simulation B. 

\begin{figure}
%\epsscale{.80}
\includegraphics{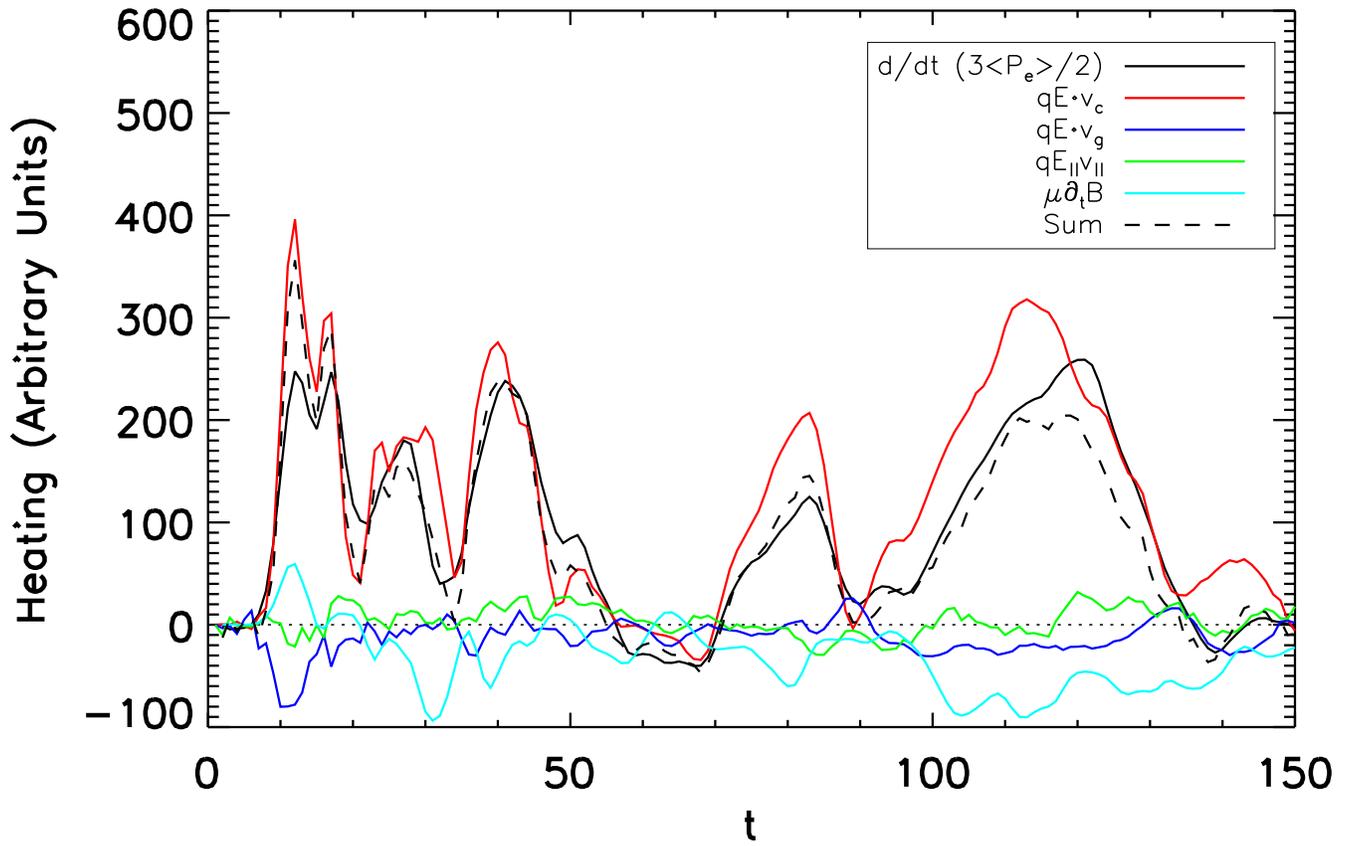}
\caption{From a simulation with a guide field of $0.2B_0$ in black the
  electron heating integrated over the upper current layer versus
  time. From Eq.~(\ref{eqn:tot_heat}) the heating from the parallel
  electric field (green), the curvature drift (red), the gradient $B$
  drift (blue), induction (cyan) and the sum (dashed black) of all of
  the heating terms in Eq.~(\ref{eqn:tot_heat}).  The curvature drift
  term, which describes Fermi reflection, dominates.}
\label{heate}
\end{figure}

\begin{figure}
%\epsscale{.80}
\includegraphics{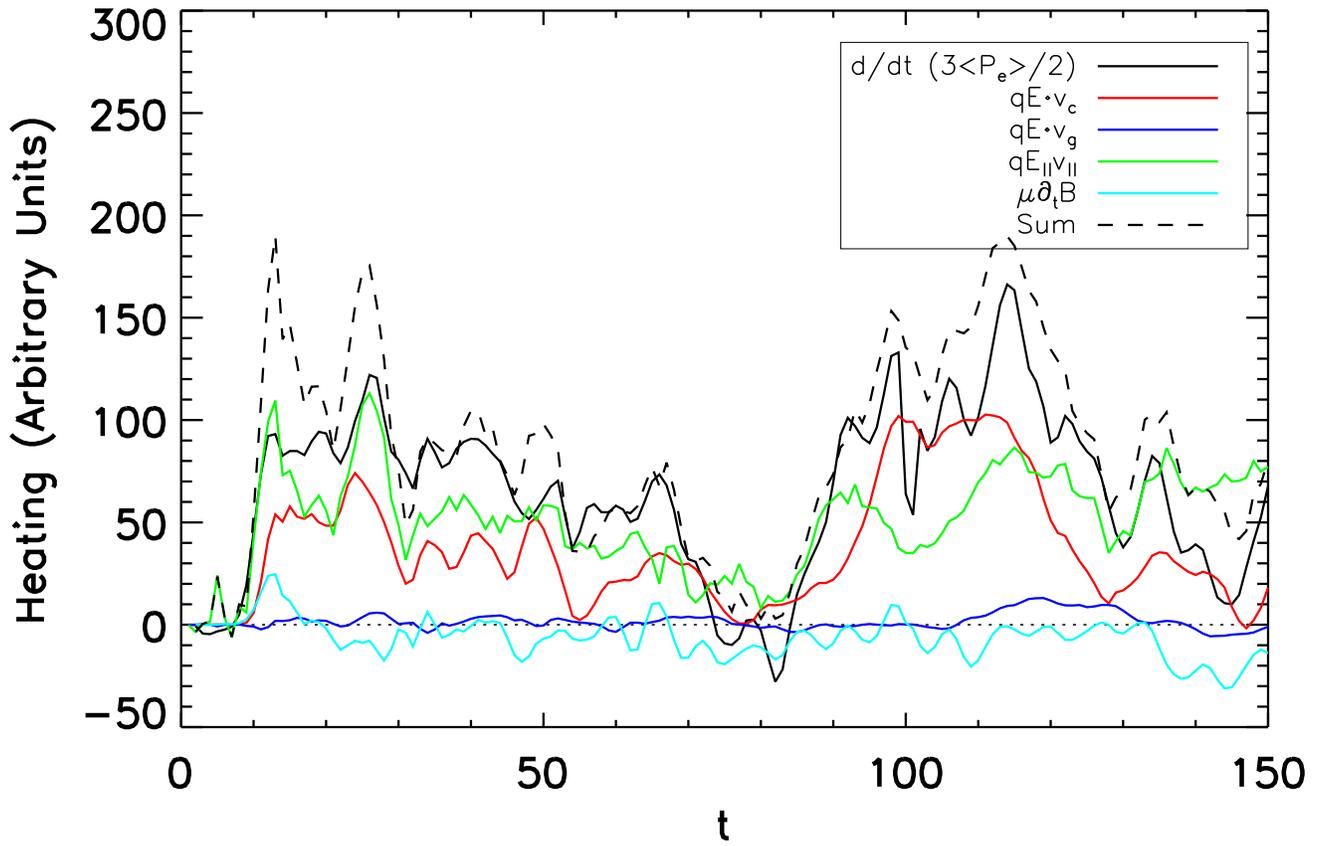}
\caption{From a simulation with a guide field of $1.0B_0$ in black the
  electron heating integrated over the upper current layer versus
  time. Other heating terms as in Fig.~\ref{heate}. In contrast with
  the case of the weak guide field in Fig.~\ref{heate}, the curvature
  and $E_\parallel$ terms are comparable in magnitude.  }
\label{heat800e}
\end{figure}

Figures \ref{heate} and \ref{heat800e} show the contributions of the
various terms in Eq.~(\ref{eqn:tot_heat}) in the upper current
sheet in simulations A and B. At a given time each term in
Eq.~(\ref{eqn:tot_heat}) was calculated at each grid point and then
integrated over space to give the displayed curves. Some smoothing was
performed to reduce the noise in the calculations but the results
shown are insensitive to its details.  The sum of heating terms on the
right-hand side of Eq.~(\ref{eqn:tot_heat}) is given by the
dashed black line, and should be compared to the solid black line
which represents the total measured electron heating. To the extent
that the two match, Eq.~(\ref{eqn:tot_heat}) represents a valid
description of the system.  The discrepancy at early time is due to
the small initial size of the islands (which makes the guiding-center
approach less accurate). Sharp, small-scale gradients that develop
during island mergers may be a source of additional discrepancies.

\begin{figure}
%\epsscale{.80}
\includegraphics{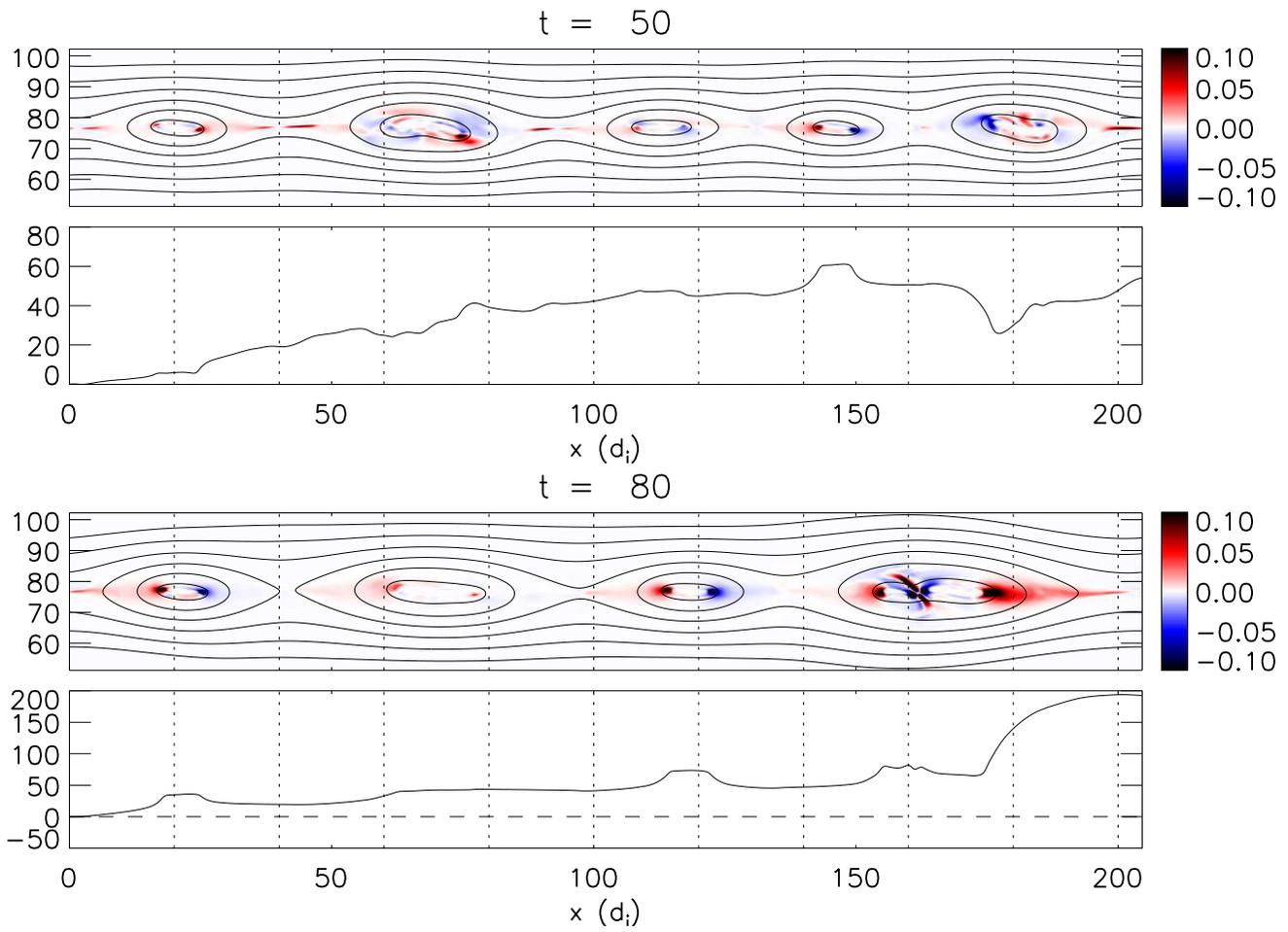}
\caption{Plots of the heating from the curvature-drift and its
  spatially integrated contribution $\Xi$ (see Eq.~(\ref{eqn:xi})) from the
  weak guide field simulation at $t = 50\Omega_{ci}^{-1}$ and $80\Omega_{ci}^{-1}$. For each time, the
  top half shows the spatial distribution and the bottom half shows
  its integrated contricution $\Xi$.  }
\label{hybrida}
\end{figure}

%Fig \ref{heate} shows the total heating for simulation A. After an
%initial period of about $20 \Omega_{ci}$ Eq. \ref{eqn:tot_heat}
%matches well with the measured heating in the system.  In
In Fig.~\ref{heate}, which corresponds to simulation A, the
curvature-drift term is the dominant source of heating and
$E_\parallel J_\parallel$ is negligible. The grad-B and $\partial_t B$
terms are also negligible and result in net cooling. This is because
magnetic reconnection releases magnetic energy and therefore reduces
the magnitude of $B$. Because of $\mu$ conservation electrons
therefore on average lose energy in the perpendicular direction. By
contrast, Fig.~\ref{heat800e} shows that in simulation B the
curvature-drift and $E_\parallel J_\parallel$ terms are comparable,
while the other terms are negligible. The increased importance of the
heating from the parallel electric field in the guide field unity case
is because of the long current layers that develop in this case
compared with those in the case of the weak guide field. Both
simulations exhibit quasi-periodic heating which is largely due to
island mergers. This can be seen by comparing times $t=50$ and $t=80$
in simulation A: the former exhibits only modest heating while the
latter exhibits strong heating. Figure \ref{hybrida} (discussed
further below) reveals that at $t=80$ two islands are merging, which
causes a burst of reconnection at the rightmost X-line in the
system. In contrast, reconnection is proceeding in the normal fashion
at $t=50$.

\begin{figure}
%\epsscale{.80}
\includegraphics{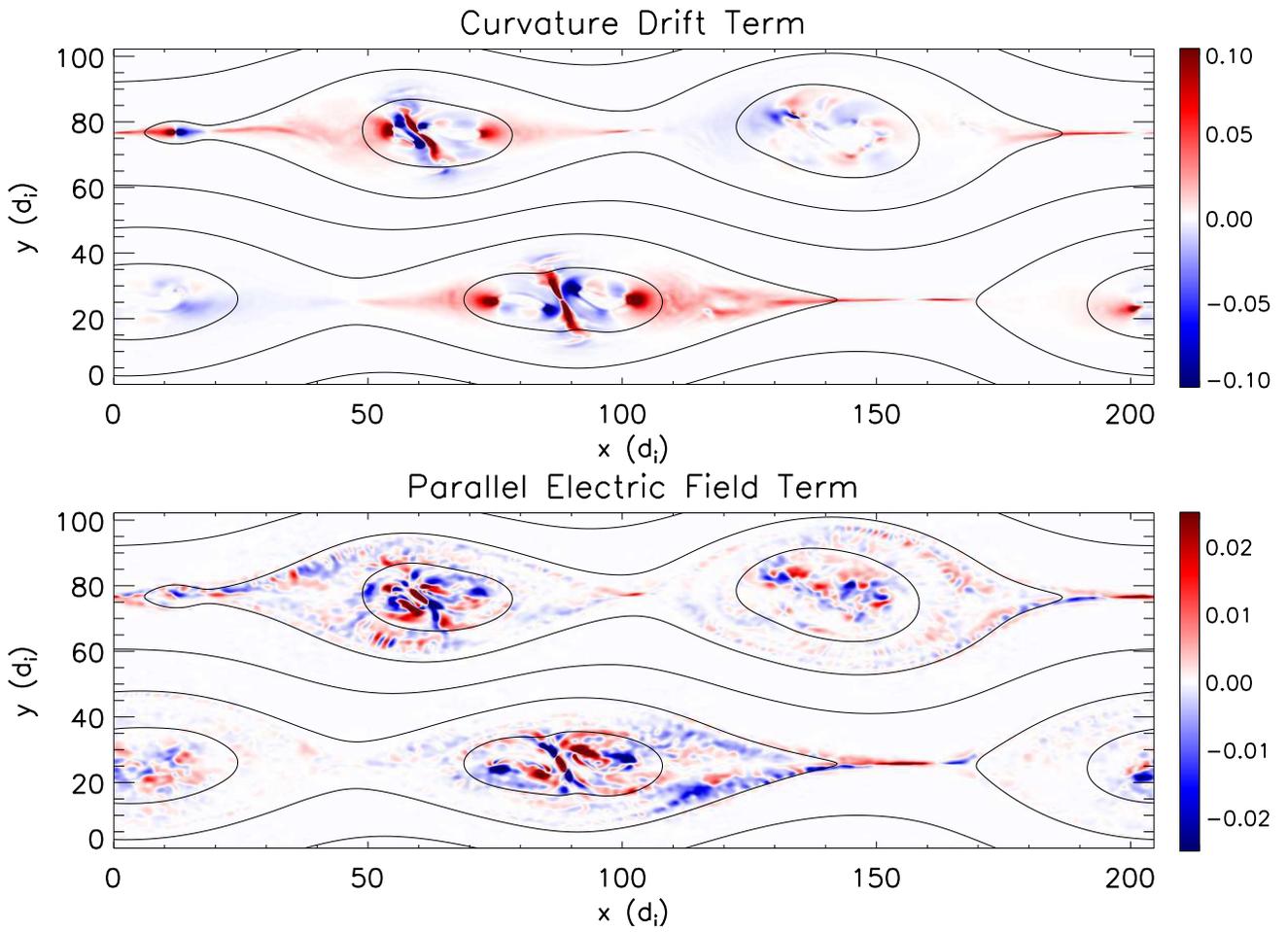}
\caption{The distribution of electron heating for a guide field of
  $0.2B_0$ at $t = 125 \Omega_{ci}^{-1}$ from the curvature (top) and
  the parallel electric field (bottom). Note the different color
  tables. The most intense heating occurs in the reconnection exhausts and
  at the ends of the islands from Fermi reflection.  }
\label{curve}
\end{figure}

\begin{figure}
%\epsscale{.80}
\includegraphics{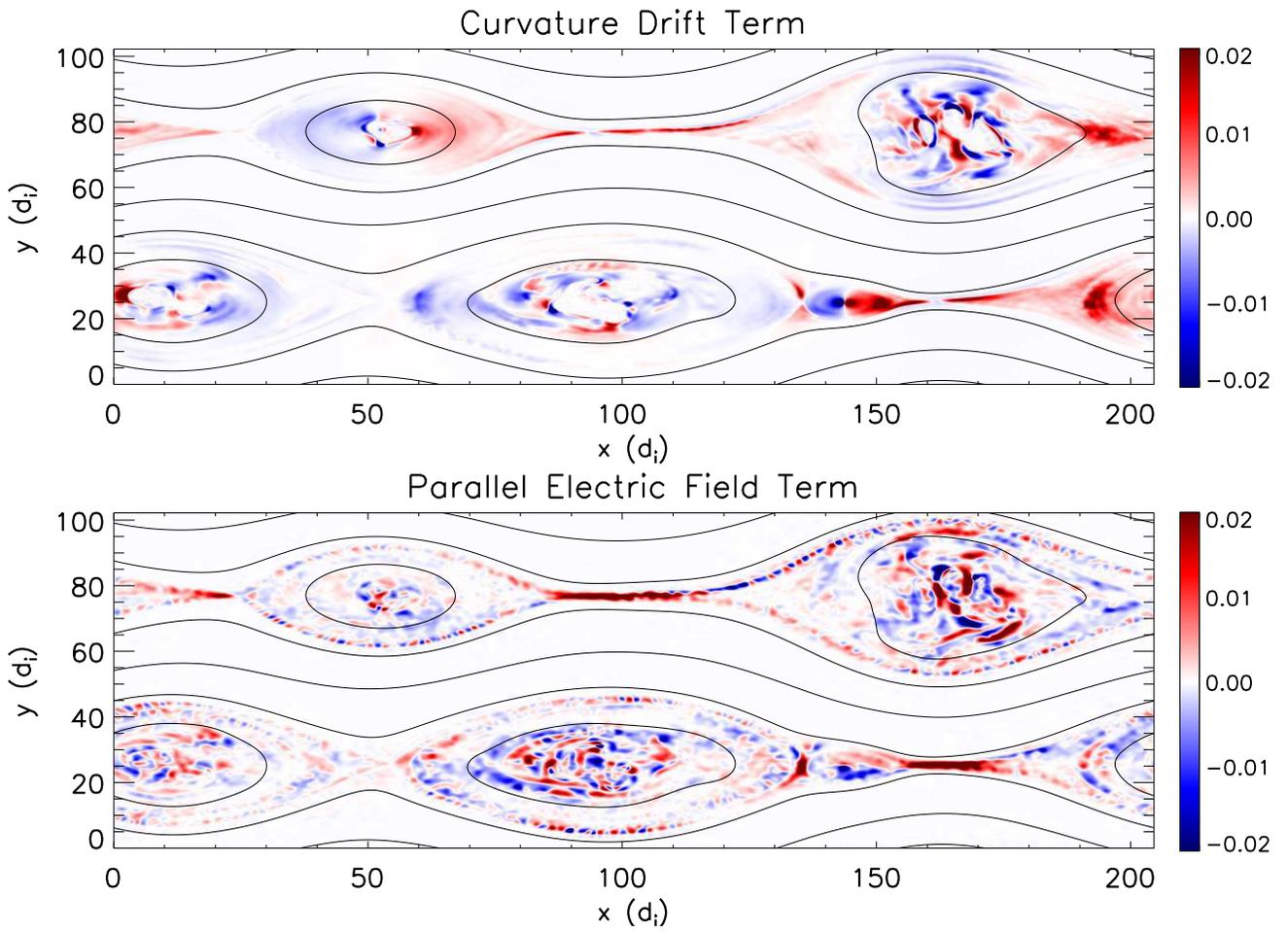}
\caption{The distribution of electron heating for a guide field of
  $1.0B_0$ at $t = 120 \Omega_{ci}^{-1}$ from the curvature (top) and
  the parallel electric field (bottom). Note that the color tables are
  the same. The current layers, where the heating from the parallel
  electric field is most intense, are much longer than in the case of
  a small guide field, }
\label{curv1e}
\end{figure}

Figures \ref{curve} and \ref{curv1e} show the spatial distribution of
the curvature and $E_\parallel J_\parallel$ terms for simulation A at
$t=120$ and B at $t=125$, respectively. As expected, the
curvature-driven heating is primarily located in the reconnection
exhaust regions and at the ends of the islands. Heating and cooling in
the island cores are due to turbulent `sloshing' of plasma inside the
island.  We show later that there is little net heating from this
behavior. The $E_\parallel J_\parallel$ term is localized near the
X-lines in both figures. The patchy regions of alternating heating and
cooling throughout the islands which is associated with electron holes
\citep{drake03a,cattell05a} does not on average produce much electron
heating (shown later). Note the different color scales in the two
plots in Fig.~\ref{curv1e}: the maximum intensity of the heating by
$E_\parallel$ is much smaller than that of the curvature drift,
consistent with its relatively small contribution to electron heating
shown in Fig.~\ref{heate}.

The patchy nature of the $E_\parallel J_\parallel$ term makes the
interpretation of this data difficult. It is not obvious, for example,
whether the heating due to $E_\parallel$ around the X-line or due to the
electrons holes dominates. As a further diagnostic, we therefore
calculate the quantity:

\begin{equation}
\label{eqn:xi}
\Xi(x) = \int^x_0 dx' \int {\mathcal U}(x',y) dy
\end{equation}
where ${\mathcal U}$ is a heating term, the $y$-integral is taken over
the half of the box containing the current layer (varying the bounds
of integration does not significantly affect the result). The slope $d
\Xi(x)/dx = \int {\mathcal U}(x,y)dy$ yields the heating at a given
$x$.

Figure \ref{hybrida} shows $\Xi$ for the curvature-drift term in
simulation A at two different times, corresponding to a temporal
minimum in the curvature-drift heating ($t = 50$) and a temporal
maximum ($t = 80$). The merger of two islands near $X \approx 160$
drives acceleration at the X-line in the far right of the
simulation. The resulting island has a larger aspect ratio, (length
$x$ compared to width $y$) so that freshly reconnected field lines
experience a greater tension force around the far right X-line. This
enhances the rate of electron heating in the exhausts around this
X-line. The plot of $\Xi$ also reveals that the heating and cooling in
island cores results in
little net heating, as can be seen for example inside the island at $x \sim
165$ at $t=80$.

\begin{figure}
%\epsscale{.80}
\includegraphics{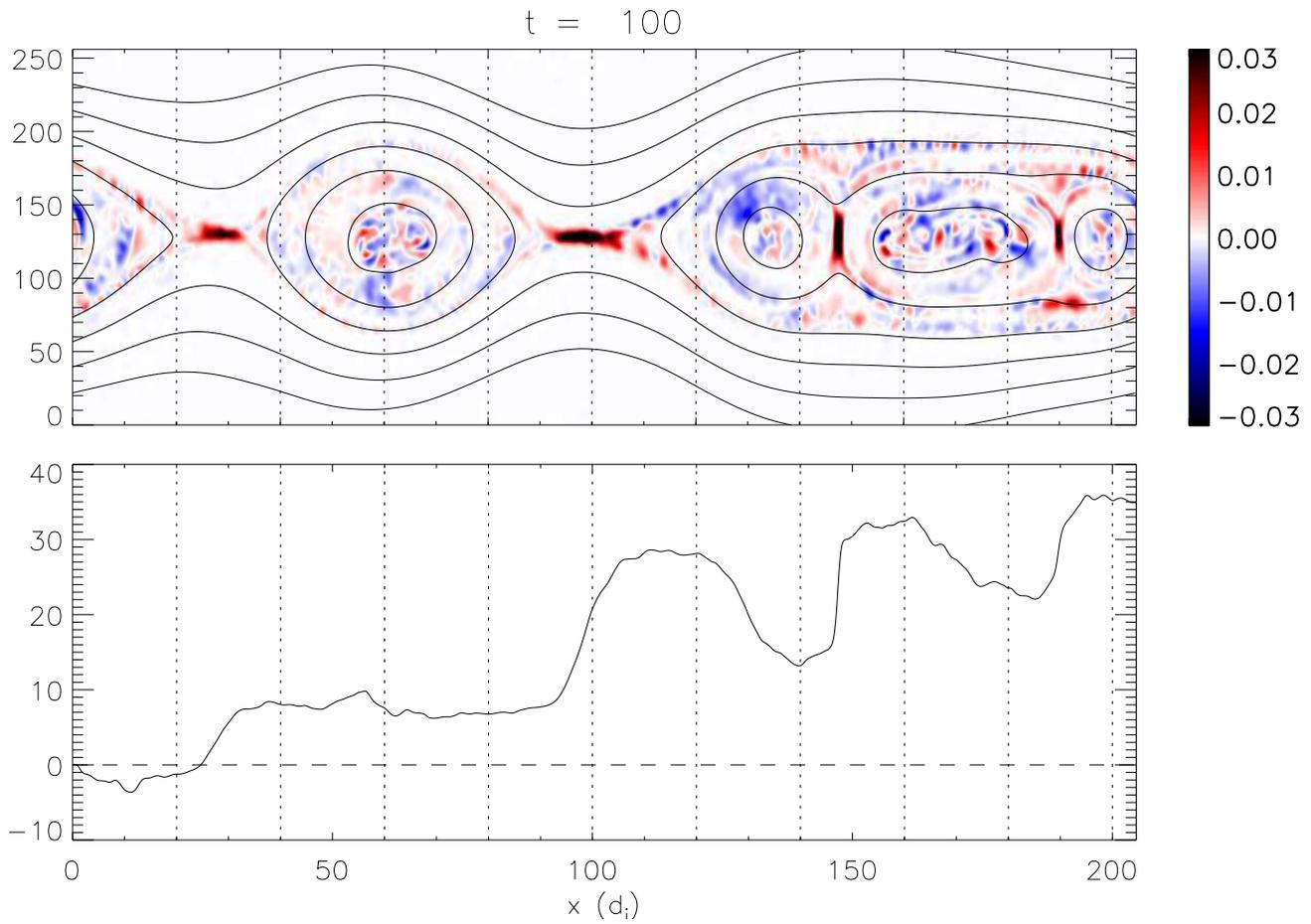}
\caption{The spatial distribution of the rate of parallel electron
  heating at $t=100\Omega_{ci}^{-1}$ from the strong guide field
  simulation (above) and its spatially integrated value $\Xi$. The
  dominant heating is from the current layers around the X-lines,
  while the contribution from electron holes in the islands appears to
  cause electron cooling.  }
\label{hybrid1e}
\end{figure}

Figure \ref{hybrid1e} shows $\Xi$ for $E_\parallel J_\parallel$ at
$t=100$ from simulation B. The
  dominant heating occurs near the primary X-lines at $x \sim 30$ and
  $100$ as well as the secondary X-lines (due to island mergers) at $x
  \sim 150$ and $190$. Inside the islands, there is net cooling. Many
  of the small scale fluctuations in the $E_\parallel J_\parallel$
  term correspond with electron holes, which are driven by electron
  beams generated near the X-line \citep{drake03a}. Because they tend
  to appear as bipolar structures in the heating term, they produce
  little net heating.

\begin{figure}
%\epsscale{.80}
\includegraphics{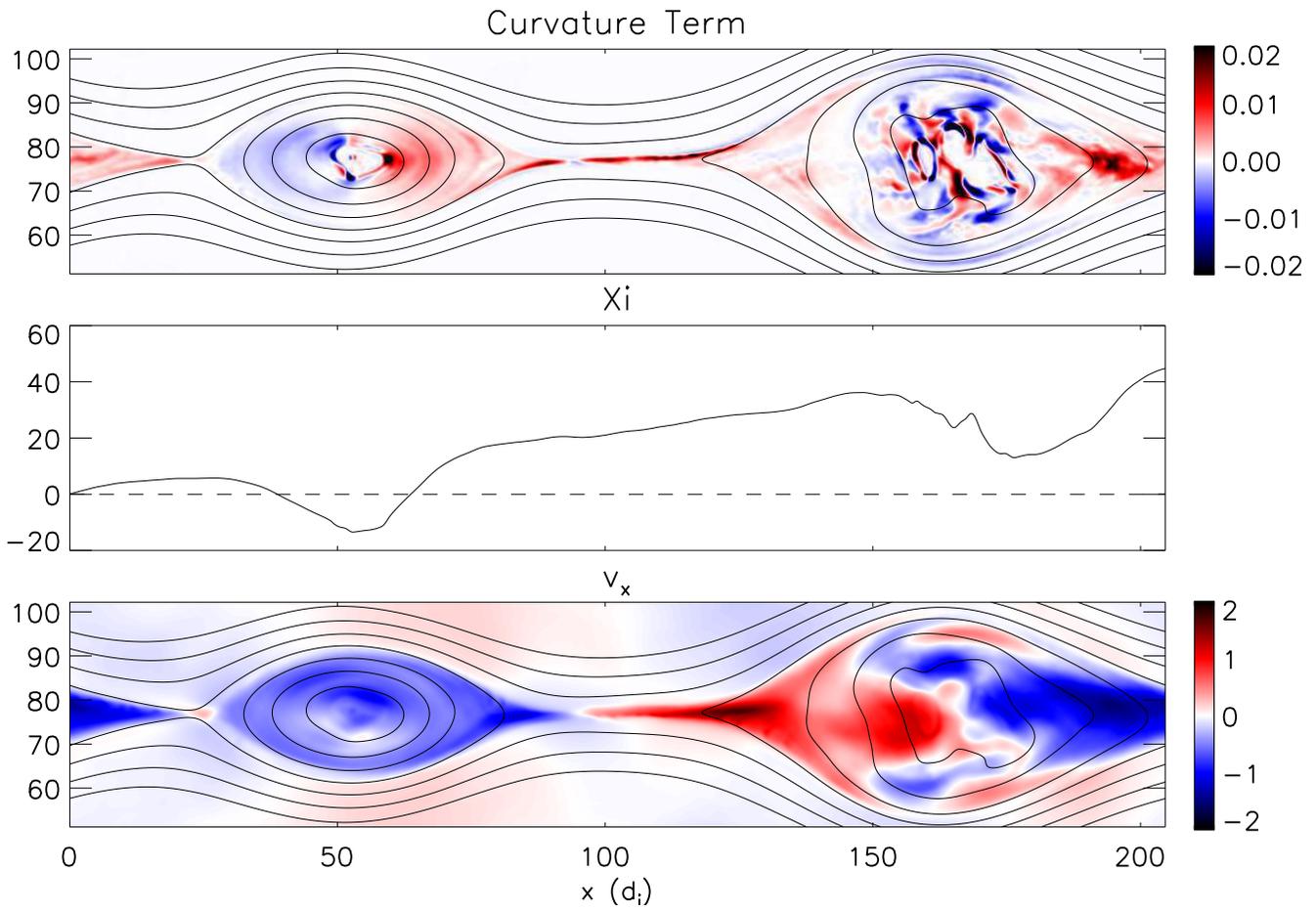}
\caption{The effect of island motion on heating from the curvature
  drift from the strong guide field simulation at
  $t=120\Omega_{ci}^{-1}$. The top panel shows the heating from the
  curvature drift the middle panel shows its spatially integrated
  contribution $\Xi$ and the bottom panel shows the horizontal bulk
  flow $v_x$.  }
\label{fig:doppler}
\end{figure}

A number of the islands exhibit dipolar heating: the curvature term
makes positive and negative contributions (red and blue) at the
opposite ends of an island. Figure \ref{fig:doppler} exhibits this
behavior. The island on the right drives heating due to Fermi
reflection at both ends, and the plot of $v_x$ shows large inward
flows indicating island contraction. By contrast, the island on the
left has dipolar heating.  The entire island is moving in the $-x$
direction. In the simulation frame, particles see receding field lines
at the left end of the island and lose energy in a
reflection. Equivalently, $\ue \cdot \boldsymbol{\kappa} < 0$.
However, the magnitude of the velocity at the right end is greater
than that at the left, so the cooling at the left end is more than
offset by the heating at the right: $\Xi$ shows that the total heating
across the island is positive.  This is ultimately an issue of
frame-dependence: in the frame of the island, both ends are
contracting towards the center so that $\ue \cdot \boldsymbol{\kappa}
> 0$.

\section{Simulation results: electron spectra}
\label{sec:spectra}

\begin{figure}
%\epsscale{.80}
\includegraphics{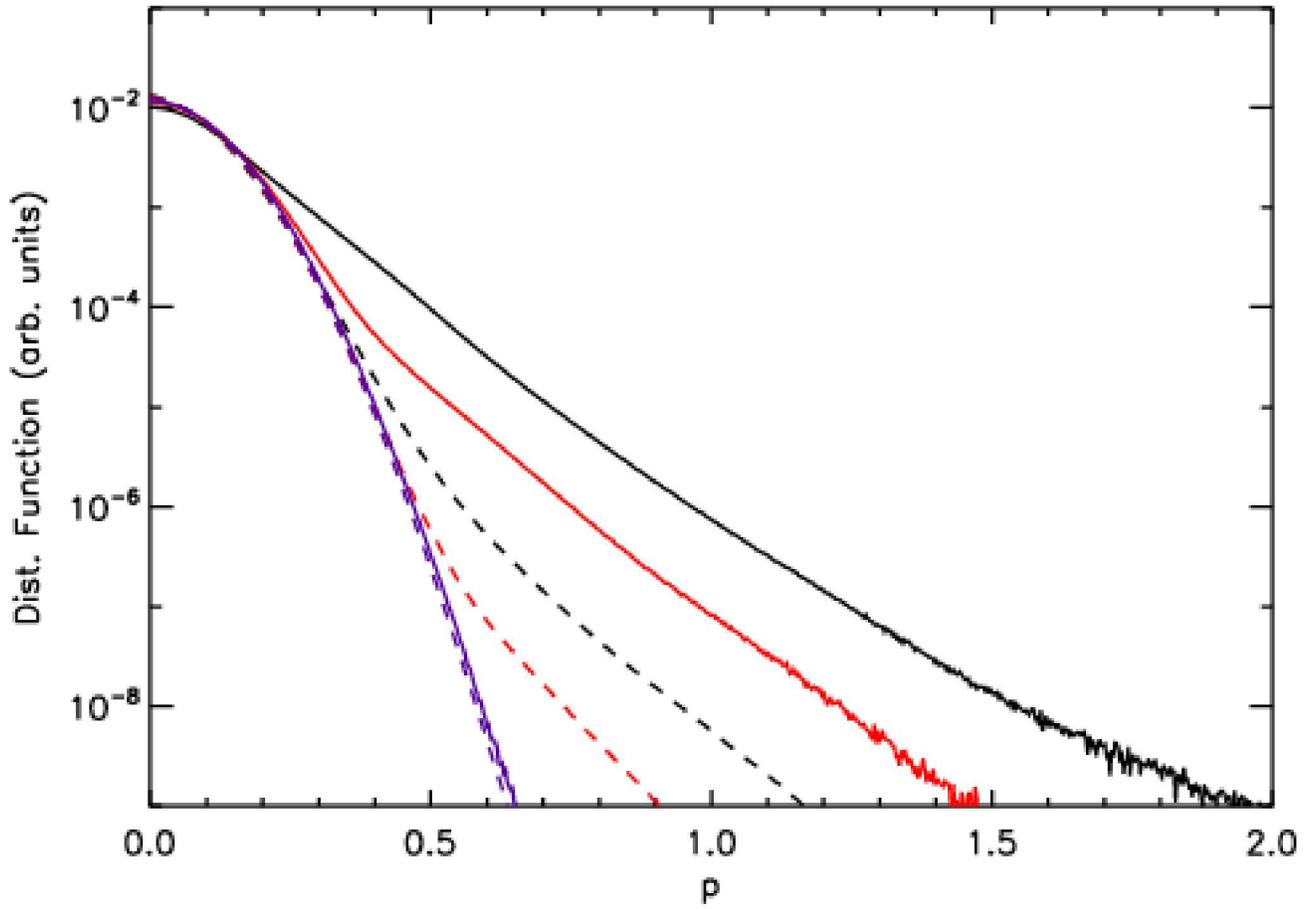}
\caption{Parallel and perpendicular electron momentum spectra (over
  the entire domain) for a simulation with guide field of $1.0B_0$ in
  a $L_x \times L_y = 819.2 \times 409.6$ domain. Solid lines
  correspond to parallel momenta and dashed with perpendicular
  momenta. Purple, red, and black are at $t=0$, $50\Omega_{ci}^{-1}$ and
  $350\Omega_{ci}^{-1}$, respectively. Note the extreme anisotropy of
  the spectra at late time. }
\label{fig:eparperp}
\end{figure}

During reconnection with a strong guide field, which is expected to be
the generic regime in most space and astrophysical systems, the
dominant mechanisms for electron acceleration are the parallel
electric field and Fermi reflection associated with the curvature
drift, both of which accelerate electrons parallel to the local
magnetic field. An important question, therefore, is whether the
energetic component of the spectrum exhibits the strong anisotropy
that is reflected in the moments $T_\parallel$ and $T_\perp$ in
Fig.~\ref{teparperp1}.  Figure \ref{fig:eparperp} shows electron
spectra for the momenta parallel and perpendicular to the magnetic
field. These spectra are taken from a simulation with the same initial
conditions as in simulation B but in a larger domain $L_x \times L_y =
819.2 \times 409.6$ carried out to $t=400$.  The larger simulation
produces much better statistics in the particle spectra compared with
simulation B shown earlier. In the parallel momentum a clear
nonthermal tail develops by $t=50$ and continues to strengthen until
the end of the simulation. The perpendicular momentum also develops a
nonthermal tail, but with an intensity that is smaller by more than
two orders of magnitude. It is hence clear that the dominant
nonthermal acceleration occurs in the parallel component and the
anisotropy survives over long periods of time as the simulation
develops. An important question is what mechanism causes the
perpendicular heating of energetic electrons. If the magnetic moment
were exactly preserved, such particles would not be produced because
the magnetic field $B$ does not reached the required values
anywhere. Therefore the increase in the perpendicular spectrum must
arise from scattering either because of non-adiabatic behavior in the
narrow boundary layers that develop as a result of reconnection or
because of the development of an instability directly driven by the
anisotropy.

\begin{figure}
%\epsscale{.80}
\includegraphics{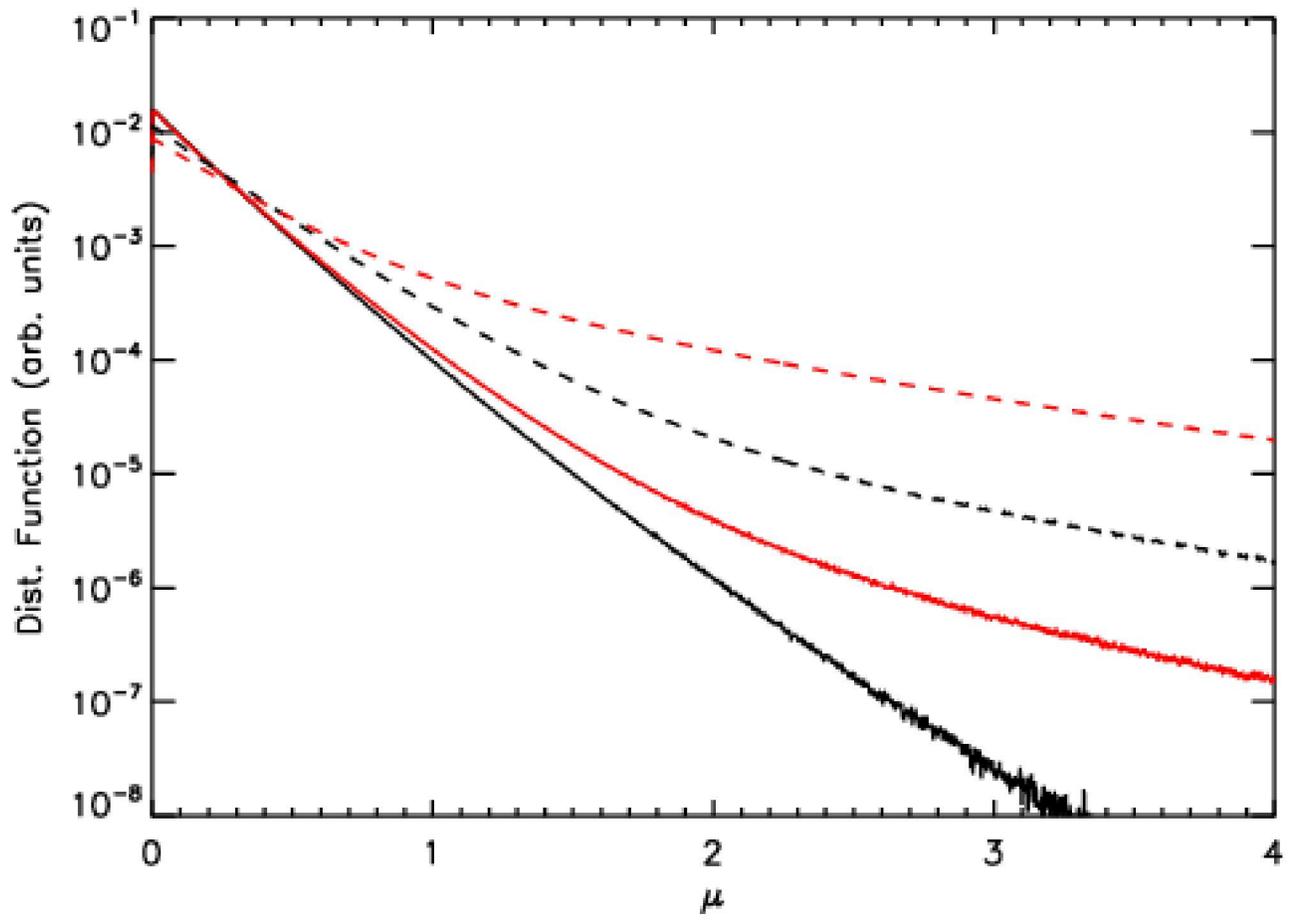}
\caption{Distribution of the electron magnetic moment $\mu = m v_\perp^2/2B$ (over the entire domain) for simulation A (dashed lines)
and B (solid lines). Black corresponds with $t=0$, red with $t=100$.
}
\label{fig:mudist}
\end{figure}

The distribution of the electron magnetic moment $\mu=mv_\perp^2/2B$ for both simulations shown in Fig.
 \ref{fig:mudist}. It is clear that $\mu$ is very well conserved in
simulation B, especially at low energies $\mu < 1$ where the
electrons remain adiabatic in the presence of the strong guide 
field $B_z = B_0$. For simulation A with $B_z = 0.2B_0$, there is a 
drop of about $10\%$ at the lowest energies, indicating that there 
is scattering into higher $\mu$. This further suggests that the 
greater perpendicular heating in simulation A is due to 
non-adiabatic behavior in the small guide field regime.

%\section{Discussion}
%\label{sec:discussion}

\section{Discussion and conclusions}
\label{sec:conclusions}

We have presented a guiding center model to explore the heating of
electrons during reconnection with modest and large guide fields. We
find that for a small guide field of $0.2B_0$ (with $B_0$ the
asymptotic reconnecting field) electron heating is dominated by the
Fermi reflection of electrons downstream of X-lines where the tension
of newly reconnected field lines drives the reconnection outflow. The
electron energy gain is given by the curvature drift of electrons in
the direction of the reconnection electric field. In this small guide
field case heating from the parallel electric field and that
associated with betatron acceleration (which is actually an energy
sink) are negligible.  In the case of a stronger guide field
($1.0B_0$) the heating associated with parallel electric fields and
the Fermi mechanism are comparable. The greater importance of the
parallel electric field is because of the elongated current layers
that form during reconnection with a guide field, which is where most
parallel heating by this mechanism takes place. The net electron
heating from electron holes, which densely populate the separatrices
and island cores, is small because positive and negative contributions
cancel.  For both weak and strong guide fields, island mergers lead to 
bursts of electron acceleration.

An important scaling question concerns the role of heating by
the parallel electric field in very large systems.  The acceleration
by parallel electric fields is largely confined to the narrow current
layers around the X-line. In contrast, the heating through Fermi
reflection occurs in a broad region in the exhaust downstream of
X-lines and well into the ends of magnetic islands. At early times the
sheer number of X-lines could well make parallel electric fields a
significant source of heating and acceleration. However, at late time
when islands may be system-size, fewer $x$-lines might remain so parallel
electric fields might not produce significant acceleration.  In
addition, the regions in which the $E_\parallel J_\parallel$ term
dominates have characteristic widths that scale with $d_e\propto
m_e^{1/2}$ with $m_e$ the electron mass. In the simulations presented
here, $m_p/m_e = 25$. For a real mass ratio of $m_p/me \approx 1836$
the corresponding regions with $E_\parallel\neq 0$ are expected to be
much smaller. In contrast, the curvature drift dominates electron
heating on island scales, which are not expected to depend on the
choice of mass ratio once islands grow to finite size.

Evidently, further simulations are required to explore how the heating
mechanisms given in Eq.~(\ref{eqn:tot_heat}) scale with system
size. One of the motivations of exploring electron acceleration in the
guiding center model is to develop a generic approach for addressing
acceleration mechanisms in 3D systems where simple explanations of
particle acceleration in contracting islands are no longer adequate:
magnetic islands will generally no longer exist because field lines in
3D systems are chaotic and therefore volume-filling. However, since
the conversion energy by the relaxation of magnetic tension is
fundamental to the reconnection process, we expect that the Fermi-like
acceleration mechanism will remain important in a 3D system and its
role can be quantified by evaluating the heating mechanisms presented
in Eq.~(\ref{eqn:tot_heat}).

Finally, we comment briefly about the implications of the strong
anisotropy of the energetic electrons seen in the spectra in
Fig.~\ref{fig:eparperp} for the simulation with a guide field of
1.0$B_0$.  Gamma-ray flares have recently been detected in the Crab
Nebula with photon energies exceeding $\approx 200$ MeV.  These
photons exceed the upper cutoff ($\approx 160$ MeV) that is obtained
by balancing energy gain from the electric field $E\sim B$ with that
from losses associated with the synchrotron radiation reaction force.
One proposed solution is that electrons are accelerated to the
necessary energies ($\approx 10^{15}$ eV) in a large-scale
reconnecting current sheet where $E\ggg B$ and the usual synchrotron
assumptions do not apply \citep{uzdensky11a}.  On the other hand,
constraining the electrons in a narrow layer and preventing their escape
into the reconnection exhaust and downstream magnetic island is a
challenge.  Another possibility is that reconnection takes place in
the presence of a guide field such that the acceleration of the
electrons is dominantly parallel to the local magnetic field so that
the anistropic energy distribution could mitigate synchrotron losses.
In such a situation a rough upper limit on reconnection-driven
energuzation can be obtained by balancing the Fermi drive (scaling as
$\gamma/R_c$, where $R_c$ is a typical radius of curvature of a
reconnecting magnetic field) against the curvature radiation loss
($\gamma^4/R_c^2$), 
\begin{equation}
\label{eqn:crab}
\gamma < (R_c/R_e)^{1/3}
\end{equation}
where $R_e = e^2/m_ec^2 = 2.82 \times 10^{-13}$ cm is the 
classical electron radius.  For the most energetic events $R_c$
should equal the system size. Based on the flare duration of 
1 day, $R_c \approx 3\times 10^{15}$ cm and the upper limit 
on the electron energy is $\epsilon = \gamma m_ec^2 \sim 10^{15}$ 
eV, which is in the range needed to explain the observations.  
Clearly, a fundamental question is whether there are 
scattering mechanisms that limit the degree of anisotropy 
of the energetic particle spectrum and therefore reduce the 
upper limit given in Eq.~(\ref{eqn:crab}).

\section{ACKNOWLEDGMENTS}
\label{sec:acknowledgements}

This work has been supported by NSF Grants AGS1202330 and PHY1102479, and NASA grants NNX11AQ93H, APL-975268, NNX08AV87G, NAS 5-98033 and NNX08AO83G. Simulations  were carried out at the National Energy Research Scientific Computing Center.

%merlin.mbs aipnum4-1.bst 2010-07-25 4.21a (PWD, AO, DPC) hacked
%Control: key (0)
%Control: author (8) initials jnrlst
%Control: editor formatted (1) identically to author
%Control: production of article title (-1) disabled
%Control: page (0) single
%Control: year (1) truncated
%Control: production of eprint (0) enabled
%

%\bibliography{paper}

\begin{thebibliography}{37}%
\makeatletter
\providecommand \@ifxundefined [1]{%
 \@ifx{#1\undefined}
}%
\providecommand \@ifnum [1]{%
 \ifnum #1\expandafter \@firstoftwo
 \else \expandafter \@secondoftwo
 \fi
}%
\providecommand \@ifx [1]{%
 \ifx #1\expandafter \@firstoftwo
 \else \expandafter \@secondoftwo
 \fi
}%
\providecommand \natexlab [1]{#1}%
\providecommand \enquote  [1]{``#1''}%
\providecommand \bibnamefont  [1]{#1}%
\providecommand \bibfnamefont [1]{#1}%
\providecommand \citenamefont [1]{#1}%
\providecommand \href@noop [0]{\@secondoftwo}%
\providecommand \href [0]{\begingroup \@sanitize@url \@href}%
\providecommand \@href[1]{\@@startlink{#1}\@@href}%
\providecommand \@@href[1]{\endgroup#1\@@endlink}%
\providecommand \@sanitize@url [0]{\catcode `\\12\catcode `\$12\catcode
  `\&12\catcode `\#12\catcode `\^12\catcode `\_12\catcode `\%12\relax}%
\providecommand \@@startlink[1]{}%
\providecommand \@@endlink[0]{}%
\providecommand \url  [0]{\begingroup\@sanitize@url \@url }%
\providecommand \@url [1]{\endgroup\@href {#1}{\urlprefix }}%
\providecommand \urlprefix  [0]{URL }%
\providecommand \Eprint [0]{\href }%
\providecommand \doibase [0]{http://dx.doi.org/}%
\providecommand \selectlanguage [0]{\@gobble}%
\providecommand \bibinfo  [0]{\@secondoftwo}%
\providecommand \bibfield  [0]{\@secondoftwo}%
\providecommand \translation [1]{[#1]}%
\providecommand \BibitemOpen [0]{}%
\providecommand \bibitemStop [0]{}%
\providecommand \bibitemNoStop [0]{.\EOS\space}%
\providecommand \EOS [0]{\spacefactor3000\relax}%
\providecommand \BibitemShut  [1]{\csname bibitem#1\endcsname}%
\let\auto@bib@innerbib\@empty
%</preamble>
\bibitem [{\citenamefont {Drenkhahn}\ and\ \citenamefont
  {Spruit}(2002)}]{drenkhahn02a}%
  \BibitemOpen
  \bibfield  {author} {\bibinfo {author} {\bibfnamefont {G.}~\bibnamefont
  {Drenkhahn}}\ and\ \bibinfo {author} {\bibfnamefont {H.~C.}\ \bibnamefont
  {Spruit}},\ }\href {\doibase 10.1051/0004-6361:20020839} {\bibfield
  {journal} {\bibinfo  {journal} {Astronomy \& Astrophysics}\ }\textbf
  {\bibinfo {volume} {391}},\ \bibinfo {pages} {1141} (\bibinfo {year}
  {2002})}\BibitemShut {NoStop}%
\bibitem [{\citenamefont {Michel}(1994)}]{michel94a}%
  \BibitemOpen
  \bibfield  {author} {\bibinfo {author} {\bibfnamefont {F.~C.}\ \bibnamefont
  {Michel}},\ }\href@noop {} {\bibfield  {journal} {\bibinfo  {journal} {Ap.
  J.}\ }\textbf {\bibinfo {volume} {431}},\ \bibinfo {pages} {397} (\bibinfo
  {year} {1994})}\BibitemShut {NoStop}%
\bibitem [{\citenamefont {Lin}\ \emph {et~al.}(2003)\citenamefont {Lin},
  \citenamefont {Krucker}, \citenamefont {Hurford}, \citenamefont {Smith},
  \citenamefont {Hudson}, \citenamefont {Holman}, \citenamefont {Schwartz},
  \citenamefont {Dennis}, \citenamefont {Share}, \citenamefont {Murphy},
  \citenamefont {Emslie}, \citenamefont {Johns-Krull},\ and\ \citenamefont
  {Vilmer}}]{lin03a}%
  \BibitemOpen
  \bibfield  {author} {\bibinfo {author} {\bibfnamefont {R.~P.}\ \bibnamefont
  {Lin}}, \bibinfo {author} {\bibfnamefont {S.}~\bibnamefont {Krucker}},
  \bibinfo {author} {\bibfnamefont {G.~J.}\ \bibnamefont {Hurford}}, \bibinfo
  {author} {\bibfnamefont {D.~M.}\ \bibnamefont {Smith}}, \bibinfo {author}
  {\bibfnamefont {H.~S.}\ \bibnamefont {Hudson}}, \bibinfo {author}
  {\bibfnamefont {G.~D.}\ \bibnamefont {Holman}}, \bibinfo {author}
  {\bibfnamefont {R.~A.}\ \bibnamefont {Schwartz}}, \bibinfo {author}
  {\bibfnamefont {B.~R.}\ \bibnamefont {Dennis}}, \bibinfo {author}
  {\bibfnamefont {G.~H.}\ \bibnamefont {Share}}, \bibinfo {author}
  {\bibfnamefont {R.~J.}\ \bibnamefont {Murphy}}, \bibinfo {author}
  {\bibfnamefont {A.~G.}\ \bibnamefont {Emslie}}, \bibinfo {author}
  {\bibfnamefont {C.}~\bibnamefont {Johns-Krull}}, \ and\ \bibinfo {author}
  {\bibfnamefont {N.}~\bibnamefont {Vilmer}},\ }\href@noop {} {\bibfield
  {journal} {\bibinfo  {journal} {Ap. J.}\ }\textbf {\bibinfo {volume} {595}},\
  \bibinfo {pages} {L69} (\bibinfo {year} {2003})}\BibitemShut {NoStop}%
\bibitem [{\citenamefont {{\O}ieroset}\ \emph {et~al.}(2002)\citenamefont
  {{\O}ieroset}, \citenamefont {Lin}, \citenamefont {Phan}, \citenamefont
  {Larson},\ and\ \citenamefont {Bale}}]{oieroset02a}%
  \BibitemOpen
  \bibfield  {author} {\bibinfo {author} {\bibfnamefont {M.}~\bibnamefont
  {{\O}ieroset}}, \bibinfo {author} {\bibfnamefont {R.~P.}\ \bibnamefont
  {Lin}}, \bibinfo {author} {\bibfnamefont {T.~D.}\ \bibnamefont {Phan}},
  \bibinfo {author} {\bibfnamefont {D.~E.}\ \bibnamefont {Larson}}, \ and\
  \bibinfo {author} {\bibfnamefont {S.~D.}\ \bibnamefont {Bale}},\ }\href@noop
  {} {\bibfield  {journal} {\bibinfo  {journal} {Phys. Rev. Lett.}\ }\textbf
  {\bibinfo {volume} {89}},\ \bibinfo {pages} {195001} (\bibinfo {year}
  {2002})}\BibitemShut {NoStop}%
\bibitem [{\citenamefont {Krucker}\ \emph {et~al.}(2010)\citenamefont
  {Krucker}, \citenamefont {Hudson}, \citenamefont {Glesener}, \citenamefont
  {White}, \citenamefont {Masuda}, \citenamefont {Wuelser},\ and\ \citenamefont
  {Lin}}]{krucker10a}%
  \BibitemOpen
  \bibfield  {author} {\bibinfo {author} {\bibfnamefont {S.}~\bibnamefont
  {Krucker}}, \bibinfo {author} {\bibfnamefont {H.~S.}\ \bibnamefont {Hudson}},
  \bibinfo {author} {\bibfnamefont {L.}~\bibnamefont {Glesener}}, \bibinfo
  {author} {\bibfnamefont {S.~M.}\ \bibnamefont {White}}, \bibinfo {author}
  {\bibfnamefont {S.}~\bibnamefont {Masuda}}, \bibinfo {author} {\bibfnamefont
  {J.-P.}\ \bibnamefont {Wuelser}}, \ and\ \bibinfo {author} {\bibfnamefont
  {R.~P.}\ \bibnamefont {Lin}},\ }\href {\doibase 10.1088/0004-637X/714/2/1108}
  {\bibfield  {journal} {\bibinfo  {journal} {Ap. J.}\ }\textbf {\bibinfo
  {volume} {714}},\ \bibinfo {pages} {1108} (\bibinfo {year}
  {2010})}\BibitemShut {NoStop}%
\bibitem [{\citenamefont {{Oka}}\ \emph {et~al.}(2013)\citenamefont {{Oka}},
  \citenamefont {{Ishikawa}}, \citenamefont {{Saint-Hilaire}}, \citenamefont
  {{Krucker}},\ and\ \citenamefont {{Lin}}}]{oka13a}%
  \BibitemOpen
  \bibfield  {author} {\bibinfo {author} {\bibfnamefont {M.}~\bibnamefont
  {{Oka}}}, \bibinfo {author} {\bibfnamefont {S.}~\bibnamefont {{Ishikawa}}},
  \bibinfo {author} {\bibfnamefont {P.}~\bibnamefont {{Saint-Hilaire}}},
  \bibinfo {author} {\bibfnamefont {S.}~\bibnamefont {{Krucker}}}, \ and\
  \bibinfo {author} {\bibfnamefont {R.~P.}\ \bibnamefont {{Lin}}},\ }\href
  {\doibase 10.1088/0004-637X/764/1/6} {\bibfield  {journal} {\bibinfo
  {journal} {Astrophys. J.}\ }\textbf {\bibinfo {volume} {764}},\ \bibinfo
  {eid} {6} (\bibinfo {year} {2013})},\ \Eprint
  {http://arxiv.org/abs/1212.2579} {arXiv:1212.2579 [astro-ph.SR]} \BibitemShut
  {NoStop}%
\bibitem [{\citenamefont {Hoshino}\ \emph {et~al.}(2001)\citenamefont
  {Hoshino}, \citenamefont {Mukai}, \citenamefont {Terasawa},\ and\
  \citenamefont {Shinohara}}]{hoshino01a}%
  \BibitemOpen
  \bibfield  {author} {\bibinfo {author} {\bibfnamefont {M.}~\bibnamefont
  {Hoshino}}, \bibinfo {author} {\bibfnamefont {T.}~\bibnamefont {Mukai}},
  \bibinfo {author} {\bibfnamefont {T.}~\bibnamefont {Terasawa}}, \ and\
  \bibinfo {author} {\bibfnamefont {I.}~\bibnamefont {Shinohara}},\ }\href@noop
  {} {\bibfield  {journal} {\bibinfo  {journal} {J. Geophys. Res.}\ }\textbf
  {\bibinfo {volume} {106}},\ \bibinfo {pages} {25,979} (\bibinfo {year}
  {2001})}\BibitemShut {NoStop}%
\bibitem [{\citenamefont {Zenitani}\ and\ \citenamefont
  {Hoshino}(2001)}]{zenitani01a}%
  \BibitemOpen
  \bibfield  {author} {\bibinfo {author} {\bibfnamefont {S.}~\bibnamefont
  {Zenitani}}\ and\ \bibinfo {author} {\bibfnamefont {M.}~\bibnamefont
  {Hoshino}},\ }\href@noop {} {\bibfield  {journal} {\bibinfo  {journal} {Ap.
  J. Lett.}\ }\textbf {\bibinfo {volume} {562}},\ \bibinfo {pages} {L63}
  (\bibinfo {year} {2001})}\BibitemShut {NoStop}%
\bibitem [{\citenamefont {Drake}\ \emph {et~al.}(2005)\citenamefont {Drake},
  \citenamefont {Shay}, \citenamefont {Thongthai},\ and\ \citenamefont
  {Swisdak}}]{drake05a}%
  \BibitemOpen
  \bibfield  {author} {\bibinfo {author} {\bibfnamefont {J.~F.}\ \bibnamefont
  {Drake}}, \bibinfo {author} {\bibfnamefont {M.~A.}\ \bibnamefont {Shay}},
  \bibinfo {author} {\bibfnamefont {W.}~\bibnamefont {Thongthai}}, \ and\
  \bibinfo {author} {\bibfnamefont {M.}~\bibnamefont {Swisdak}},\ }\href@noop
  {} {\bibfield  {journal} {\bibinfo  {journal} {Phys. Rev. Lett.}\ }\textbf
  {\bibinfo {volume} {94}},\ \bibinfo {pages} {095001} (\bibinfo {year}
  {2005})}\BibitemShut {NoStop}%
\bibitem [{\citenamefont {Pritchett}(2006)}]{pritchett06a}%
  \BibitemOpen
  \bibfield  {author} {\bibinfo {author} {\bibfnamefont {P.~L.}\ \bibnamefont
  {Pritchett}},\ }\href {\doibase 10.1029/2006JA011793} {\bibfield  {journal}
  {\bibinfo  {journal} {J. Geophys. Res.}\ }\textbf {\bibinfo {volume} {111}},\
  \bibinfo {eid} {A10212} (\bibinfo {year} {2006}),\
  10.1029/2006JA011793}\BibitemShut {NoStop}%
\bibitem [{\citenamefont {Egedal}\ \emph {et~al.}(2009)\citenamefont {Egedal},
  \citenamefont {Daughton}, \citenamefont {Drake}, \citenamefont {Katz},\ and\
  \citenamefont {L{\^e}}}]{egedal09a}%
  \BibitemOpen
  \bibfield  {author} {\bibinfo {author} {\bibfnamefont {J.}~\bibnamefont
  {Egedal}}, \bibinfo {author} {\bibfnamefont {W.}~\bibnamefont {Daughton}},
  \bibinfo {author} {\bibfnamefont {J.~F.}\ \bibnamefont {Drake}}, \bibinfo
  {author} {\bibfnamefont {N.}~\bibnamefont {Katz}}, \ and\ \bibinfo {author}
  {\bibfnamefont {A.}~\bibnamefont {L{\^e}}},\ }\href {\doibase
  10.1063/1.3130732} {\bibfield  {journal} {\bibinfo  {journal} {Phys.
  Plasmas}\ }\textbf {\bibinfo {volume} {16}},\ \bibinfo {eid} {050701}
  (\bibinfo {year} {2009}),\ 10.1063/1.3130732}\BibitemShut {NoStop}%
\bibitem [{\citenamefont {Oka}\ \emph {et~al.}(2010)\citenamefont {Oka},
  \citenamefont {Phan}, \citenamefont {Krucker}, \citenamefont {Fujimoto},\
  and\ \citenamefont {Shinohara}}]{oka10a}%
  \BibitemOpen
  \bibfield  {author} {\bibinfo {author} {\bibfnamefont {M.}~\bibnamefont
  {Oka}}, \bibinfo {author} {\bibfnamefont {T.-D.}\ \bibnamefont {Phan}},
  \bibinfo {author} {\bibfnamefont {S.}~\bibnamefont {Krucker}}, \bibinfo
  {author} {\bibfnamefont {M.}~\bibnamefont {Fujimoto}}, \ and\ \bibinfo
  {author} {\bibfnamefont {I.}~\bibnamefont {Shinohara}},\ }\href {\doibase
  10.1088/0004-637X/714/1/915} {\bibfield  {journal} {\bibinfo  {journal} {Ap.
  J.}\ }\textbf {\bibinfo {volume} {714}},\ \bibinfo {pages} {915} (\bibinfo
  {year} {2010})}\BibitemShut {NoStop}%
\bibitem [{\citenamefont {Litvinenko}(1996)}]{litvinenko96a}%
  \BibitemOpen
  \bibfield  {author} {\bibinfo {author} {\bibfnamefont {Y.~E.}\ \bibnamefont
  {Litvinenko}},\ }\href@noop {} {\bibfield  {journal} {\bibinfo  {journal}
  {Ap. J.}\ }\textbf {\bibinfo {volume} {462}},\ \bibinfo {pages} {997}
  (\bibinfo {year} {1996})}\BibitemShut {NoStop}%
\bibitem [{\citenamefont {Egedal}, \citenamefont {Daughton},\ and\
  \citenamefont {L{\^e}}(2012)}]{egedal12a}%
  \BibitemOpen
  \bibfield  {author} {\bibinfo {author} {\bibfnamefont {J.}~\bibnamefont
  {Egedal}}, \bibinfo {author} {\bibfnamefont {W.}~\bibnamefont {Daughton}}, \
  and\ \bibinfo {author} {\bibfnamefont {A.}~\bibnamefont {L{\^e}}},\ }\href
  {\doibase 10.1038/nphys2249} {\bibfield  {journal} {\bibinfo  {journal}
  {Nature Phys.}\ }\textbf {\bibinfo {volume} {8}},\ \bibinfo {pages} {321}
  (\bibinfo {year} {2012})}\BibitemShut {NoStop}%
\bibitem [{\citenamefont {Drake}\ \emph
  {et~al.}(2006{\natexlab{a}})\citenamefont {Drake}, \citenamefont {Swisdak},
  \citenamefont {Che},\ and\ \citenamefont {Shay}}]{drake06a}%
  \BibitemOpen
  \bibfield  {author} {\bibinfo {author} {\bibfnamefont {J.~F.}\ \bibnamefont
  {Drake}}, \bibinfo {author} {\bibfnamefont {M.}~\bibnamefont {Swisdak}},
  \bibinfo {author} {\bibfnamefont {H.}~\bibnamefont {Che}}, \ and\ \bibinfo
  {author} {\bibfnamefont {M.~A.}\ \bibnamefont {Shay}},\ }\href {\doibase
  10.1038/nature05116} {\bibfield  {journal} {\bibinfo  {journal} {Nature}\
  }\textbf {\bibinfo {volume} {443}},\ \bibinfo {pages} {553} (\bibinfo {year}
  {2006}{\natexlab{a}})}\BibitemShut {NoStop}%
\bibitem [{\citenamefont {Drake}\ \emph {et~al.}(2010)\citenamefont {Drake},
  \citenamefont {Opher}, \citenamefont {Swisdak},\ and\ \citenamefont
  {Chamoun}}]{drake10a}%
  \BibitemOpen
  \bibfield  {author} {\bibinfo {author} {\bibfnamefont {J.~F.}\ \bibnamefont
  {Drake}}, \bibinfo {author} {\bibfnamefont {M.}~\bibnamefont {Opher}},
  \bibinfo {author} {\bibfnamefont {M.}~\bibnamefont {Swisdak}}, \ and\
  \bibinfo {author} {\bibfnamefont {J.~N.}\ \bibnamefont {Chamoun}},\ }\href
  {\doibase 10.1088/0004-637X/709/2/963} {\bibfield  {journal} {\bibinfo
  {journal} {Ap. J.}\ }\textbf {\bibinfo {volume} {709}},\ \bibinfo {pages}
  {963} (\bibinfo {year} {2010})}\BibitemShut {NoStop}%
\bibitem [{\citenamefont {Drake}\ \emph
  {et~al.}(2006{\natexlab{b}})\citenamefont {Drake}, \citenamefont {Swisdak},
  \citenamefont {Schoeffler}, \citenamefont {Rogers},\ and\ \citenamefont
  {Kobayashi}}]{drake06b}%
  \BibitemOpen
  \bibfield  {author} {\bibinfo {author} {\bibfnamefont {J.~F.}\ \bibnamefont
  {Drake}}, \bibinfo {author} {\bibfnamefont {M.}~\bibnamefont {Swisdak}},
  \bibinfo {author} {\bibfnamefont {K.~M.}\ \bibnamefont {Schoeffler}},
  \bibinfo {author} {\bibfnamefont {B.~N.}\ \bibnamefont {Rogers}}, \ and\
  \bibinfo {author} {\bibfnamefont {S.}~\bibnamefont {Kobayashi}},\ }\href
  {\doibase 10.1029/2006GL025957} {\bibfield  {journal} {\bibinfo  {journal}
  {Geophys. Res. Lett.}\ }\textbf {\bibinfo {volume} {33}},\ \bibinfo {eid}
  {L13105} (\bibinfo {year} {2006}{\natexlab{b}}),\
  10.1029/2006GL025957}\BibitemShut {NoStop}%
\bibitem [{\citenamefont {Daughton}\ \emph {et~al.}(2011)\citenamefont
  {Daughton}, \citenamefont {Roytershteyn}, \citenamefont {Karimabadi},
  \citenamefont {Yin}, \citenamefont {Albright}, \citenamefont {Bergen},\ and\
  \citenamefont {Bowers}}]{daughton11a}%
  \BibitemOpen
  \bibfield  {author} {\bibinfo {author} {\bibfnamefont {W.}~\bibnamefont
  {Daughton}}, \bibinfo {author} {\bibfnamefont {V.}~\bibnamefont
  {Roytershteyn}}, \bibinfo {author} {\bibfnamefont {H.}~\bibnamefont
  {Karimabadi}}, \bibinfo {author} {\bibfnamefont {L.}~\bibnamefont {Yin}},
  \bibinfo {author} {\bibfnamefont {B.~J.}\ \bibnamefont {Albright}}, \bibinfo
  {author} {\bibfnamefont {B.}~\bibnamefont {Bergen}}, \ and\ \bibinfo {author}
  {\bibfnamefont {K.~J.}\ \bibnamefont {Bowers}},\ }\href {\doibase
  10.1038/nphys1965} {\bibfield  {journal} {\bibinfo  {journal} {Nature Phys.}\
  }\textbf {\bibinfo {volume} {7}},\ \bibinfo {pages} {539} (\bibinfo {year}
  {2011})}\BibitemShut {NoStop}%
\bibitem [{\citenamefont {Biskamp}(1986)}]{biskamp86a}%
  \BibitemOpen
  \bibfield  {author} {\bibinfo {author} {\bibfnamefont {D.}~\bibnamefont
  {Biskamp}},\ }\href@noop {} {\bibfield  {journal} {\bibinfo  {journal} {Phys.
  Fluids}\ }\textbf {\bibinfo {volume} {29}},\ \bibinfo {pages} {1520}
  (\bibinfo {year} {1986})}\BibitemShut {NoStop}%
\bibitem [{\citenamefont {Loureiro}, \citenamefont {Schekochihin},\ and\
  \citenamefont {Cowley}(2007)}]{loureiro07a}%
  \BibitemOpen
  \bibfield  {author} {\bibinfo {author} {\bibfnamefont {N.~F.}\ \bibnamefont
  {Loureiro}}, \bibinfo {author} {\bibfnamefont {A.~A.}\ \bibnamefont
  {Schekochihin}}, \ and\ \bibinfo {author} {\bibfnamefont {S.~C.}\
  \bibnamefont {Cowley}},\ }\href {\doibase 10.1063/1.2783986} {\bibfield
  {journal} {\bibinfo  {journal} {Phys. Plasmas}\ }\textbf {\bibinfo {volume}
  {14}},\ \bibinfo {eid} {100703} (\bibinfo {year} {2007}),\
  10.1063/1.2783986}\BibitemShut {NoStop}%
\bibitem [{\citenamefont {Cassak}, \citenamefont {Shay},\ and\ \citenamefont
  {Drake}(2009)}]{cassak09a}%
  \BibitemOpen
  \bibfield  {author} {\bibinfo {author} {\bibfnamefont {P.~A.}\ \bibnamefont
  {Cassak}}, \bibinfo {author} {\bibfnamefont {M.~A.}\ \bibnamefont {Shay}}, \
  and\ \bibinfo {author} {\bibfnamefont {J.~F.}\ \bibnamefont {Drake}},\ }\href
  {\doibase 10.1063/1.3274462} {\bibfield  {journal} {\bibinfo  {journal}
  {Phys. Plasmas}\ }\textbf {\bibinfo {volume} {16}},\ \bibinfo {eid} {120702}
  (\bibinfo {year} {2009}),\ 10.1063/1.3274462}\BibitemShut {NoStop}%
\bibitem [{\citenamefont {Huang}\ and\ \citenamefont
  {Bhattacharjee}(2010)}]{huang10a}%
  \BibitemOpen
  \bibfield  {author} {\bibinfo {author} {\bibfnamefont {Y.-M.}\ \bibnamefont
  {Huang}}\ and\ \bibinfo {author} {\bibfnamefont {A.}~\bibnamefont
  {Bhattacharjee}},\ }\href {\doibase 10.1063/1.3420208} {\bibfield  {journal}
  {\bibinfo  {journal} {Phys. Plasmas}\ }\textbf {\bibinfo {volume} {17}},\
  \bibinfo {eid} {062104} (\bibinfo {year} {2010}),\
  10.1063/1.3420208}\BibitemShut {NoStop}%
\bibitem [{\citenamefont {Dmitruk}, \citenamefont {Matthaeus},\ and\
  \citenamefont {Seenu}(2004)}]{dmitruk04a}%
  \BibitemOpen
  \bibfield  {author} {\bibinfo {author} {\bibfnamefont {P.}~\bibnamefont
  {Dmitruk}}, \bibinfo {author} {\bibfnamefont {W.~H.}\ \bibnamefont
  {Matthaeus}}, \ and\ \bibinfo {author} {\bibfnamefont {N.}~\bibnamefont
  {Seenu}},\ }\href {\doibase 10.1086/425301} {\bibfield  {journal} {\bibinfo
  {journal} {Ap. J.}\ }\textbf {\bibinfo {volume} {617}},\ \bibinfo {pages}
  {667} (\bibinfo {year} {2004})}\BibitemShut {NoStop}%
\bibitem [{\citenamefont {Onofri}, \citenamefont {Isliker},\ and\ \citenamefont
  {Vlahos}(2006)}]{onofri06a}%
  \BibitemOpen
  \bibfield  {author} {\bibinfo {author} {\bibfnamefont {M.}~\bibnamefont
  {Onofri}}, \bibinfo {author} {\bibfnamefont {H.}~\bibnamefont {Isliker}}, \
  and\ \bibinfo {author} {\bibfnamefont {L.}~\bibnamefont {Vlahos}},\ }\href
  {\doibase 10.1103/PhysRevLett96.151102} {\bibfield  {journal} {\bibinfo
  {journal} {Phys. Rev. Lett.}\ }\textbf {\bibinfo {volume} {96}},\ \bibinfo
  {eid} {151102} (\bibinfo {year} {2006}),\
  10.1103/PhysRevLett96.151102}\BibitemShut {NoStop}%
\bibitem [{\citenamefont {Swisdak}\ \emph {et~al.}(2005)\citenamefont
  {Swisdak}, \citenamefont {Drake}, \citenamefont {McIlhargey},\ and\
  \citenamefont {Shay}}]{swisdak05a}%
  \BibitemOpen
  \bibfield  {author} {\bibinfo {author} {\bibfnamefont {M.}~\bibnamefont
  {Swisdak}}, \bibinfo {author} {\bibfnamefont {J.~F.}\ \bibnamefont {Drake}},
  \bibinfo {author} {\bibfnamefont {J.}~\bibnamefont {McIlhargey}}, \ and\
  \bibinfo {author} {\bibfnamefont {M.~A.}\ \bibnamefont {Shay}},\ }\href
  {\doibase 10.1029/2004JA010748} {\bibfield  {journal} {\bibinfo  {journal}
  {J. Geophys. Res.}\ }\textbf {\bibinfo {volume} {110}},\ \bibinfo {eid}
  {A05210} (\bibinfo {year} {2005}),\ 10.1029/2004JA010748}\BibitemShut
  {NoStop}%
\bibitem [{\citenamefont {Knizhnik}, \citenamefont {Swisdak},\ and\
  \citenamefont {Drake}(2011)}]{knizhnik11a}%
  \BibitemOpen
  \bibfield  {author} {\bibinfo {author} {\bibfnamefont {K.}~\bibnamefont
  {Knizhnik}}, \bibinfo {author} {\bibfnamefont {M.}~\bibnamefont {Swisdak}}, \
  and\ \bibinfo {author} {\bibfnamefont {J.}~\bibnamefont {Drake}},\ }\href
  {\doibase 10.1088/2041-8205/743/2/L35} {\bibfield  {journal} {\bibinfo
  {journal} {Ap. J. Lett.}\ }\textbf {\bibinfo {volume} {743}} (\bibinfo {year}
  {2011}),\ 10.1088/2041-8205/743/2/L35}\BibitemShut {NoStop}%
\bibitem [{\citenamefont {Drake}\ and\ \citenamefont
  {Swisdak}(2014)}]{drake14a}%
  \BibitemOpen
  \bibfield  {author} {\bibinfo {author} {\bibfnamefont {J.~F.}\ \bibnamefont
  {Drake}}\ and\ \bibinfo {author} {\bibfnamefont {M.}~\bibnamefont
  {Swisdak}},\ }\href@noop {} {\bibfield  {journal} {\bibinfo  {journal} {Phys.
  Plasmas}\ } (\bibinfo {year} {2014})},\ \bibinfo {note} {submitted,
  arXiv:1404.7795}\BibitemShut {NoStop}%
\bibitem [{\citenamefont {Guo}\ \emph {et~al.}(2014)\citenamefont {Guo},
  \citenamefont {Li}, \citenamefont {Daughton},\ and\ \citenamefont
  {Liu}}]{guo14a}%
  \BibitemOpen
  \bibfield  {author} {\bibinfo {author} {\bibfnamefont {F.}~\bibnamefont
  {Guo}}, \bibinfo {author} {\bibfnamefont {H.}~\bibnamefont {Li}}, \bibinfo
  {author} {\bibfnamefont {W.}~\bibnamefont {Daughton}}, \ and\ \bibinfo
  {author} {\bibfnamefont {Y.-H.}\ \bibnamefont {Liu}},\ }\href@noop {}
  {\bibfield  {journal} {\bibinfo  {journal} {Phys. Rev. Lett.}\ } (\bibinfo
  {year} {2014})},\ \bibinfo {note} {submitted, arXiv:1405.4040}\BibitemShut
  {NoStop}%
\bibitem [{\citenamefont {Northrop}(1963)}]{northrop63a}%
  \BibitemOpen
  \bibfield  {author} {\bibinfo {author} {\bibfnamefont {T.~G.}\ \bibnamefont
  {Northrop}},\ }\enquote {\bibinfo {title} {The adiabatic motion of charged
  particles},}\ \ (\bibinfo  {publisher} {Interscience Publishers},\ \bibinfo
  {year} {1963})\BibitemShut {NoStop}%
\bibitem [{\citenamefont {Hoshino}, \citenamefont {Mukai},\ and\ \citenamefont
  {Kokubun}(1998)}]{hoshino98a}%
  \BibitemOpen
  \bibfield  {author} {\bibinfo {author} {\bibfnamefont {M.}~\bibnamefont
  {Hoshino}}, \bibinfo {author} {\bibfnamefont {T.}~\bibnamefont {Mukai}}, \
  and\ \bibinfo {author} {\bibfnamefont {S.}~\bibnamefont {Kokubun}},\
  }\href@noop {} {\bibfield  {journal} {\bibinfo  {journal} {J. Geophys. Res.}\
  }\textbf {\bibinfo {volume} {103}},\ \bibinfo {pages} {4509} (\bibinfo {year}
  {1998})}\BibitemShut {NoStop}%
\bibitem [{\citenamefont {Gosling}\ \emph {et~al.}(2005)\citenamefont
  {Gosling}, \citenamefont {Skoug}, \citenamefont {McComas},\ and\
  \citenamefont {Smith}}]{gosling05a}%
  \BibitemOpen
  \bibfield  {author} {\bibinfo {author} {\bibfnamefont {J.~T.}\ \bibnamefont
  {Gosling}}, \bibinfo {author} {\bibfnamefont {R.~M.}\ \bibnamefont {Skoug}},
  \bibinfo {author} {\bibfnamefont {D.~J.}\ \bibnamefont {McComas}}, \ and\
  \bibinfo {author} {\bibfnamefont {C.~W.}\ \bibnamefont {Smith}},\ }\href
  {\doibase 10.1029/2004JA010809} {\bibfield  {journal} {\bibinfo  {journal}
  {J. Geophys. Res.}\ }\textbf {\bibinfo {volume} {110}},\ \bibinfo {eid}
  {A01107} (\bibinfo {year} {2005}),\ 10.1029/2004JA010809}\BibitemShut
  {NoStop}%
\bibitem [{\citenamefont {Phan}\ \emph {et~al.}(2007)\citenamefont {Phan},
  \citenamefont {Paschmann}, \citenamefont {Twitty}, \citenamefont {Mozer},
  \citenamefont {Gosling}, \citenamefont {Eastwood}, \citenamefont
  {{\O}ieroset}, \citenamefont {R\`eme},\ and\ \citenamefont
  {Lucek}}]{phan07a}%
  \BibitemOpen
  \bibfield  {author} {\bibinfo {author} {\bibfnamefont {T.~D.}\ \bibnamefont
  {Phan}}, \bibinfo {author} {\bibfnamefont {G.}~\bibnamefont {Paschmann}},
  \bibinfo {author} {\bibfnamefont {C.}~\bibnamefont {Twitty}}, \bibinfo
  {author} {\bibfnamefont {F.~S.}\ \bibnamefont {Mozer}}, \bibinfo {author}
  {\bibfnamefont {J.~T.}\ \bibnamefont {Gosling}}, \bibinfo {author}
  {\bibfnamefont {J.~P.}\ \bibnamefont {Eastwood}}, \bibinfo {author}
  {\bibfnamefont {M.}~\bibnamefont {{\O}ieroset}}, \bibinfo {author}
  {\bibfnamefont {H.}~\bibnamefont {R\`eme}}, \ and\ \bibinfo {author}
  {\bibfnamefont {E.~A.}\ \bibnamefont {Lucek}},\ }\href {\doibase
  10.1029/2007GL030343} {\bibfield  {journal} {\bibinfo  {journal} {Geophys.
  Res. Lett.}\ ,\ \bibinfo {eid} {L14104}} (\bibinfo {year} {2007}),\
  10.1029/2007GL030343}\BibitemShut {NoStop}%
\bibitem [{\citenamefont {Zeiler}\ \emph {et~al.}(2002)\citenamefont {Zeiler},
  \citenamefont {Biskamp}, \citenamefont {Drake}, \citenamefont {Rogers},
  \citenamefont {Shay},\ and\ \citenamefont {Scholer}}]{zeiler02a}%
  \BibitemOpen
  \bibfield  {author} {\bibinfo {author} {\bibfnamefont {A.}~\bibnamefont
  {Zeiler}}, \bibinfo {author} {\bibfnamefont {D.}~\bibnamefont {Biskamp}},
  \bibinfo {author} {\bibfnamefont {J.~F.}\ \bibnamefont {Drake}}, \bibinfo
  {author} {\bibfnamefont {B.~N.}\ \bibnamefont {Rogers}}, \bibinfo {author}
  {\bibfnamefont {M.~A.}\ \bibnamefont {Shay}}, \ and\ \bibinfo {author}
  {\bibfnamefont {M.}~\bibnamefont {Scholer}},\ }\href {\doibase
  10.1029/2001JA000287} {\bibfield  {journal} {\bibinfo  {journal} {J. Geophys.
  Res.}\ }\textbf {\bibinfo {volume} {107}},\ \bibinfo {pages} {1230} (\bibinfo
  {year} {2002})}\BibitemShut {NoStop}%
\bibitem [{\citenamefont {Harris}(1962)}]{harris62a}%
  \BibitemOpen
  \bibfield  {author} {\bibinfo {author} {\bibfnamefont {E.~G.}\ \bibnamefont
  {Harris}},\ }\href {\doibase 10.1007/BF02733547} {\bibfield  {journal}
  {\bibinfo  {journal} {Nuovo Cim.}\ }\textbf {\bibinfo {volume} {23}},\
  \bibinfo {pages} {115} (\bibinfo {year} {1962})}\BibitemShut {NoStop}%
\bibitem [{\citenamefont {Drake}\ \emph {et~al.}(2003)\citenamefont {Drake},
  \citenamefont {Swisdak}, \citenamefont {Cattell}, \citenamefont {Shay},
  \citenamefont {Rogers},\ and\ \citenamefont {Zeiler}}]{drake03a}%
  \BibitemOpen
  \bibfield  {author} {\bibinfo {author} {\bibfnamefont {J.~F.}\ \bibnamefont
  {Drake}}, \bibinfo {author} {\bibfnamefont {M.}~\bibnamefont {Swisdak}},
  \bibinfo {author} {\bibfnamefont {C.}~\bibnamefont {Cattell}}, \bibinfo
  {author} {\bibfnamefont {M.~A.}\ \bibnamefont {Shay}}, \bibinfo {author}
  {\bibfnamefont {B.~N.}\ \bibnamefont {Rogers}}, \ and\ \bibinfo {author}
  {\bibfnamefont {A.}~\bibnamefont {Zeiler}},\ }\href@noop {} {\bibfield
  {journal} {\bibinfo  {journal} {Science}\ }\textbf {\bibinfo {volume}
  {299}},\ \bibinfo {pages} {873} (\bibinfo {year} {2003})}\BibitemShut
  {NoStop}%
\bibitem [{\citenamefont {Cattell}\ \emph {et~al.}(2005)\citenamefont
  {Cattell}, \citenamefont {Dombeck}, \citenamefont {Wygant}, \citenamefont
  {Drake}, \citenamefont {Swisdak}, \citenamefont {Goldstein}, \citenamefont
  {Keith}, \citenamefont {Fazakerley}, \citenamefont {Andr\'{e}}, \citenamefont
  {Lucek},\ and\ \citenamefont {Balogh}}]{cattell05a}%
  \BibitemOpen
  \bibfield  {author} {\bibinfo {author} {\bibfnamefont {C.}~\bibnamefont
  {Cattell}}, \bibinfo {author} {\bibfnamefont {J.}~\bibnamefont {Dombeck}},
  \bibinfo {author} {\bibfnamefont {J.}~\bibnamefont {Wygant}}, \bibinfo
  {author} {\bibfnamefont {J.~F.}\ \bibnamefont {Drake}}, \bibinfo {author}
  {\bibfnamefont {M.}~\bibnamefont {Swisdak}}, \bibinfo {author} {\bibfnamefont
  {M.~L.}\ \bibnamefont {Goldstein}}, \bibinfo {author} {\bibfnamefont
  {W.}~\bibnamefont {Keith}}, \bibinfo {author} {\bibfnamefont
  {A.}~\bibnamefont {Fazakerley}}, \bibinfo {author} {\bibfnamefont
  {M.}~\bibnamefont {Andr\'{e}}}, \bibinfo {author} {\bibfnamefont
  {E.}~\bibnamefont {Lucek}}, \ and\ \bibinfo {author} {\bibfnamefont
  {A.}~\bibnamefont {Balogh}},\ }\href {\doibase 10.1029/2004JA010519}
  {\bibfield  {journal} {\bibinfo  {journal} {J. Geophys. Res.}\ }\textbf
  {\bibinfo {volume} {110}},\ \bibinfo {eid} {A01211} (\bibinfo {year}
  {2005}),\ 10.1029/2004JA010519}\BibitemShut {NoStop}%
\bibitem [{\citenamefont {Uzdensky}, \citenamefont {Cerutti},\ and\
  \citenamefont {Begelman}(2011)}]{uzdensky11a}%
  \BibitemOpen
  \bibfield  {author} {\bibinfo {author} {\bibfnamefont {D.~A.}\ \bibnamefont
  {Uzdensky}}, \bibinfo {author} {\bibfnamefont {B.}~\bibnamefont {Cerutti}}, \
  and\ \bibinfo {author} {\bibfnamefont {M.~C.}\ \bibnamefont {Begelman}},\
  }\href@noop {} {\bibfield  {journal} {\bibinfo  {journal} {The Astrophysical
  Journal Letters}\ }\textbf {\bibinfo {volume} {737}},\ \bibinfo {pages} {L40}
  (\bibinfo {year} {2011})}\BibitemShut {NoStop}%
\end{thebibliography}

\clearpage

\clearpage

%\begin{figure}
%\epsscale{.80}
%\includegraphics{p14fermitraj.pdf}
%\caption{
%}
%\label{ferCumulative integral of $E_\parallel J_\parallel$.
%Most of the heating occurs near the X-lines.mitraj}
%\end{figure}
%\clearpage

\clearpage

\clearpage

\clearpage

\clearpage

\clearpage

\clearpage

\clearpage

\clearpage

\clearpage

\clearpage

\clearpage

\clearpage

\end{document}